\documentclass[usenatbib]{mn2e}
\usepackage{graphicx}
\usepackage{ifthen}
\usepackage{url}
\usepackage{fixltx2e}

\def\ltsima{$\; \buildrel < \over \sim \;$}
\def\lta{\lower.5ex\hbox{\ltsima}}
\def\gtsima{$\; \buildrel > \over \sim \;$}
\def\simgt{\lower.5ex\hbox{\gtsima}}
\def\kms{{\rm\,km\,s^{-1}}}

\def\kpc{{\rm\,kpc}}

\def\msun{{\rm\,M_\odot}}
\def\lsun{{\rm\,L_\odot}}

\def\AA{$\; \buildrel \circ \over {\mathrm A}$}

\def\deg{^\circ}
\def\s{\ifmmode \widetilde \else \~\fi}
\def\={\overline}

\def\spose#1{\hbox to 0pt{#1\hss}}

\def\lta{\mathrel{\spose{\lower 3pt\hbox{$\mathchar"218$}}
     \raise 2.0pt\hbox{$\mathchar"13C$}}}
\def\gta{\mathrel{\spose{\lower 3pt\hbox{$\mathchar"218$}}
     \raise 2.0pt\hbox{$\mathchar"13E$}}}
\def\Dt{\spose{\raise 1.5ex\hbox{\hskip3pt$\mathchar"201$}}}    % upper case
\def\dt{\spose{\raise 1.0ex\hbox{\hskip2pt$\mathchar"201$}}}    % lower case

\def\dotsfill{\leaders\hbox to 1em{\hss.\hss}\hfill}

\def\feh{{\rm[Fe/H]}}

\long\def\symbolfootnote[#1]#2{\begingroup%
\def\thefootnote{\fnsymbol{footnote}}\footnote[#1]{#2}\endgroup}

\def\ec4{EC4}

\def\andv{And\,V}

\loadboldmathitalic 

\title[]{The scatter about the ``Universal'' dwarf spheroidal mass
  profile: A kinematic study of the M31 satellites, And V and And VI}
\author [M.\ L.\ M.\ Collins et al.] {M. L. M. Collins$^{1,*}$,
  S. C. Chapman$^1$, R. M.  Rich$^2$, M. J. Irwin$^1$,
  J. Pe\~narrubia$^1$,\newauthor R. A. Ibata$^3$, N. Arimoto$^4$,
  A. M. Brooks$^5$, A. M. N. Ferguson$^6$, G. F. Lewis$^7$,
  \newauthor A. W. McConnachie$^8$, K. Venn$^{9}$\\ $^1$Institute of
  Astronomy,Madingley Rise, Cambridge, CB3 0HA ,UK\\ $^2$Department of
  Physics and Astronomy, University of California, Los Angeles, CA
  90095-1547 \\ $^3$Observatoire de Strasbourg,11, rue de
  l'Universit\'e, F-67000, Strasbourg, France\\ $^4$National
  Astronomical Observatory of Japan, Osawa 2-21-1, Mitaka, Tokyo,
  Japan\\ $^5$California Institute of Technology, M/C 350-17,
  Pasadena, CA 91125, USA\\ $^6$ Institute for Astronomy, University
  of Edinburgh, Royal Observatory, Blackford Hill, Edinburgh, UK EH9
  3HJ\\ $^7$Sydney Institute for Astronomy, School of Physics, A29,
  University of Sydney, NSW 2006, Australia \\ $^8$NRC Herzberg
  Institute for Astrophysics, 5071 West Saanich Road, Victoria,
  British Columbia, Canada, V9E 2E7 \\ $^{9}$Dept. of Physics \&
  Astronomy, University of Victoria, 3800 Finerty Road, Victoria, BC
  V8P 1A1, Canada\\ } \date{\today}
\begin{document}
\maketitle 

\begin{abstract}

While the satellites of the Milky Way (MW) have been shown to be
largely consistent in terms of their mass contained within one
half--light radius ($M_{half}$) with a ``universal'' mass profile, a
number of M31 satellites are found to be inconsistent with these
relations, and seem kinematically colder in their central regions than
their MW cousins. In this work, we present new kinematic and updated
structural properties for two M31 dSphs, And V and And VI using data
from the Keck Low Resolution Imaging Spectrograph (LRIS) and the DEep
Imaging Multi-Object Spectrograph (DEIMOS) instruments and the Subaru
Suprime-Cam imager. We measure systemic velocities of
$v_r=-393.1\pm4.2\kms$ and $-344.8\pm2.5\kms$, and dispersions of
$\sigma_v=11.5^{+5.3}_{-4.4}\kms$ and $\sigma_v=9.4^{+3.2}_{-2.4}\kms$
for And V and And VI respectively, meaning these two objects are
consistent with the trends in $\sigma_v$ and $r_{half}$ set by their
MW counterparts. We also investigate the nature of this scatter about
the MW dSph mass profiles for the ``Classical'' (i.e. $M_V<-8$) MW and
M31 dSphs. When comparing both the ``classical'' MW and M31 dSphs to
the best--fit mass profiles in the size--velocity dispersion plane, we
find general scatter in both the positive (i.e. hotter) and negative
(i.e. colder) directions from these profiles. However, barring one
exception (CVnI) only the M31 dSphs are found to scatter towards a
colder regime, and, excepting the And I dSph, only MW objects scatter
to hotter dispersions. The scatter for the combined
population is greater than expected from measurement errors alone. We
assess this divide in the context of the differing disc-to-halo mass
(i.e. stars and baryons to total virial mass) ratios of the two hosts
and argue that the underlying mass profiles for dSphs differ from
galaxy to galaxy, and are modified by the baryonic component of the
host.

\end{abstract}

\section{Introduction}

\symbolfootnote[2]{$^*$email: mlmc2@ast.cam.ac.uk}

The past few years have been a revelation for the kinematic properties
of the dwarf spheroidals (dSphs) of the Local Group.
\citet{strigari08} compiled kinematic data for 18 of the Milky Way
dSph galaxies using a maximum likelihood technique based on the Jeans
equation to determine the masses for each of these systems within a
300 pc radius (M$_{300}$). This radius was chosen as the masses of
these objects are best constrained within the region where there are
tracers of the potential (i.e. stars) and for their sample, 300 pc
represented the average radius for this region. They determined that
despite a luminous range of more than 4 orders of magnitude, the
objects were consistent with having a dynamical mass of 10$^7\msun$
within 300 pc of their centre, and declared this as a common mass
scale for dSph galaxies. This characteristic mass scale had already
been observed for the brighter dSphs ($M_V\le-8$, \citealt{mateo98}),
but the consistency of the fainter objects was a surprise. Further,
they showed that the dSphs were all consistent with having formed in
stellar halos with total masses $\gta10^9\msun$, which could implicate
this as the cut-off mass for star formation within cold dark matter
halos, or the minimum mass with which a dark matter halo could
form. This worked was then extended by \citet{walker09b} and
\citet{wolf10}, both of whom show that the Milky Way (MW) dSph halos
exhibit a correlation between global velocity dispersion and
half-light radius, and as such, the central densities of their dark
matter halos do not show significant scatter over a large range of
luminous scale-radii. \citet{walker09b} used this observation to
postulate that the dSphs of the MW appear to be consistent with having
formed with a ``Universal'' dark matter halo mass profile. In
addition, follow-up work by \citet{walker10} demonstrated that the
halos of these dark matter dominated systems are also consistent with
the mean rotation curve derived for spiral galaxies in
\citet{mcgaugh07}, indicating a constant dark matter central surface
density for galaxies ranging from MW dSphs to spirals.
Recent results presented by both the Pan-Andromeda Archaeological
Survey (PAndAS) and the Spectroscopic and Photometric Landscape of
Andromeda's Stellar Halo (SPLASH) groups have indicated significant
differences between the kinematic temperatures (or velocity
dispersions) of the dark matter halos of dSphs orbiting the MW
vs. those orbiting Andromeda, with a number of the latter being
seemingly colder for a given half-light radius, and therefore less
massive and less dense \citep{collins10,kalirai10}, signalling that
this ``Universal'' mass profile may not extend to dSphs found outside
of the MW. They are also outliers to the \citet{mcgaugh07} rotation
curve relations as discussed in \citet{walker10}. This strongly
suggests that while many dSphs appear similar in terms of their
central densities, they are not all embedded within dark matter halos
that follow a ``Universal'' profile.

Interpreting the observed differences in the subhalos for these two
populations is not trivial as it requires an understanding of why some
dSphs are born, or currently reside in, dark matter halos with
differing physical and dynamical properties. One explanation could be
that the underlying physical processes behind the formation of dSph
halos differs between the MW and M31; this clearly confronts our
current cosmological paradigm for the growth of dark matter
halos. Another possibility is that the M31 subhalos could have formed
later than the MW population when the Universe was less dense,
resulting in colder, less dense subhalos. The MW is thought to have
had a more quiescent merger history than M31, accreting less baryons
over cosmic time. This means that the MW could have `formed' earlier
(i.e. reached its half mass earlier) than M31. It is unclear how much
later it would need to form to reproduce these results so more
modelling is required to investigate this scenario fully. Finally,
recent work by \citet{penarrubia10} put forward a more physically
motivated theory for this difference. Using N-body simulations, they
demonstrated that subhalos that evolve in host environments with
disc-to-halo mass ratios of twice that measured in the MW naturally
end up with less mass within their half-light radii, very similar to
the observations of the M31 population. This fits nicely with the
results of \citet{hammer07}, where they report that M31 has been more
successful at accreting baryons than the MW, resulting in a stellar
population that is $\sim2.5$ times the mass of the MW stellar
population.

While there is a tendency for M31 dSphs to inhabit colder, lower mass
halos, it is by no means observed throughout the population. Almost
half of all spectroscopically surveyed M31 satellites (5 out of 12 -- And I,
And VII, And XIII, And XV and And XVI
\citealt{letarte09,collins10,kalirai10}) are observed to have velocity
dispersions that are entirely consistent with their MW counterparts,
suggesting an overall scatter within this population. Therefore, how
significant is this subset of kinematically colder M31 dSphs? Could
this trend be the result of observational biases present in both
systems? Due to our position in the MW, located within the disc at a
distance of 785 kpc from M31, we are obviously not able to observe
exactly the same subset of dSph galaxies in our own halo compared to
that of M31. In the MW, we are hampered by obscuration from components
of our own Galaxy that prevent an areal coverage of the
system. However, for the region that has been covered in large all-sky
surveys (such as SDSS) we are thought to be complete in our detections
of bright, ``classical''-type dSphs. \citep{tollerud08,koposov08} In
M31 the limiting factors in dwarf detection are attributed to the
large distance between ourselves and our neighbour, which prevents us
from observing objects with $M_V>-6$. We also struggle to identifying
dwarf galaxies that lie close to the centre of M31 (at
$R_{M31}<40\kpc$) as these objects are obscured (either directly or
via projection effects) by the large disc component of the galaxy
(traced out as far as 40 kpc kinematically and 25--30 kpc
photometrically). We also observe more bright dSphs at large distances
from the centre of M31 than we do in the MW, resulting in a mean
distance of satellite from host of 184 kpc in M31 compared with 138
kpc in the MW \citep{mcconnachie09,richardson11}.

The study of the intrinsic properties of dSphs has also been
illuminated by the surprising behaviour that has been observed in some
of the least luminous members of this population, the ``ultra-faint''
dSphs. Since 2005, a large number of very faint objects ($-2>M_V>-7$,
e.g. \citealt{willman05b,belokurov06b,belokurov07,zucker06a,zucker06b,belokurov08})
have been discovered, and have revealed several unusual properties for
the population. For example, they appear to be hugely dark matter
dominated, resulting in a significant departure from the
well-established mass-luminosity relationship for such objects
\citep{mateo98}. The exact morphological nature of these objects is
also a subject of some controversy as they possess properties that are
common to both dSphs and globular clusters and are therefore difficult
to classify as either, such as the unusual objects, Willman 1 and
Segue 1
\citep{willman05a,belokurov07,niederste09,simon10,willman10}. They may
also be far from dynamical equilibrium, and instead are undergoing
extreme tidal disruption, inflating their observed velocity
dispersions and inferred mass-to-light ratios (e.g. Segue 1 and
Hercules, \citealt{niederste09,sand09} and \citealt{jin10}), resulting
in a misleading representation of the behaviour of these
low-luminosity satellites. These faint objects are observed within a
regime where inflation of the velocity dispersion from binary star
systems is non-negligible \citep{mcconnachie10,minor10}, meaning their
dynamical dispersions may be lower than the values quoted in the
literature, drastically altering the inferred $M/L$ ratios. Therefore,
if we restrict ourselves to comparing solely the dSphs whose kinematic
properties are free from such uncertainties for the two host galaxies,
does this variation persist? And what of its significance?

To answer these questions, we have embarked upon a limited comparison
of ``classical'' M31 and MW dSphs, where we define ``classical'' to be
any object with $M_V<-7.9$, with particular focus on two M31 dSphs
that have recently been spectroscopically observed, And V and And VI,
allowing us to present their kinematic properties for the first
time. As And V and And VI represent the last two classical dSphs that
belong to either the MW or M31 to have their velocity dispersions and
masses measured, these observations make for a timely consideration of
the whole ensemble of classical dSphs.

\section{Observations and analysis techniques}

\begin{figure*}
\begin{center}
\includegraphics[angle=0,width=0.4\hsize]{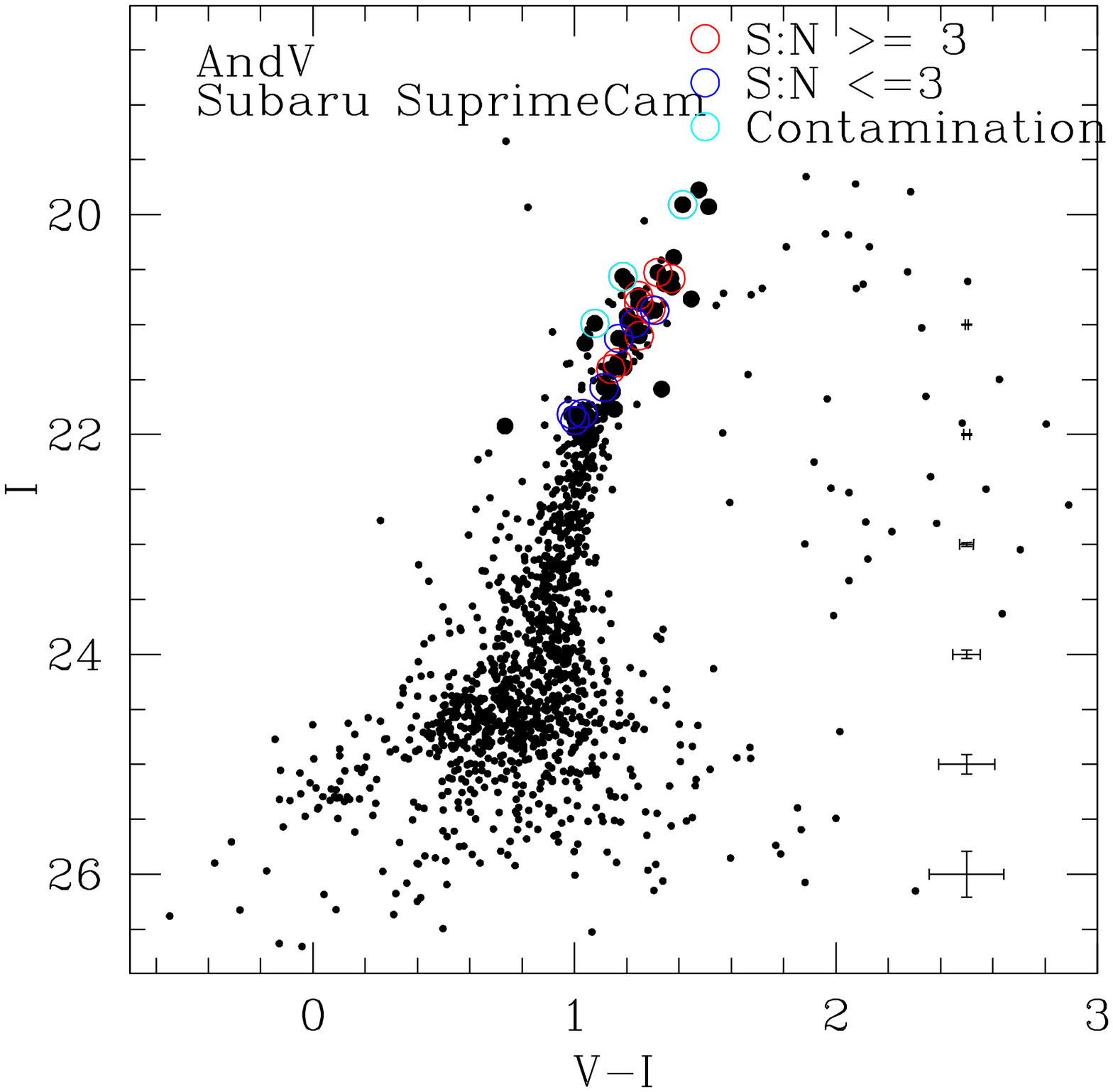}
\includegraphics[angle=0,width=0.4\hsize]{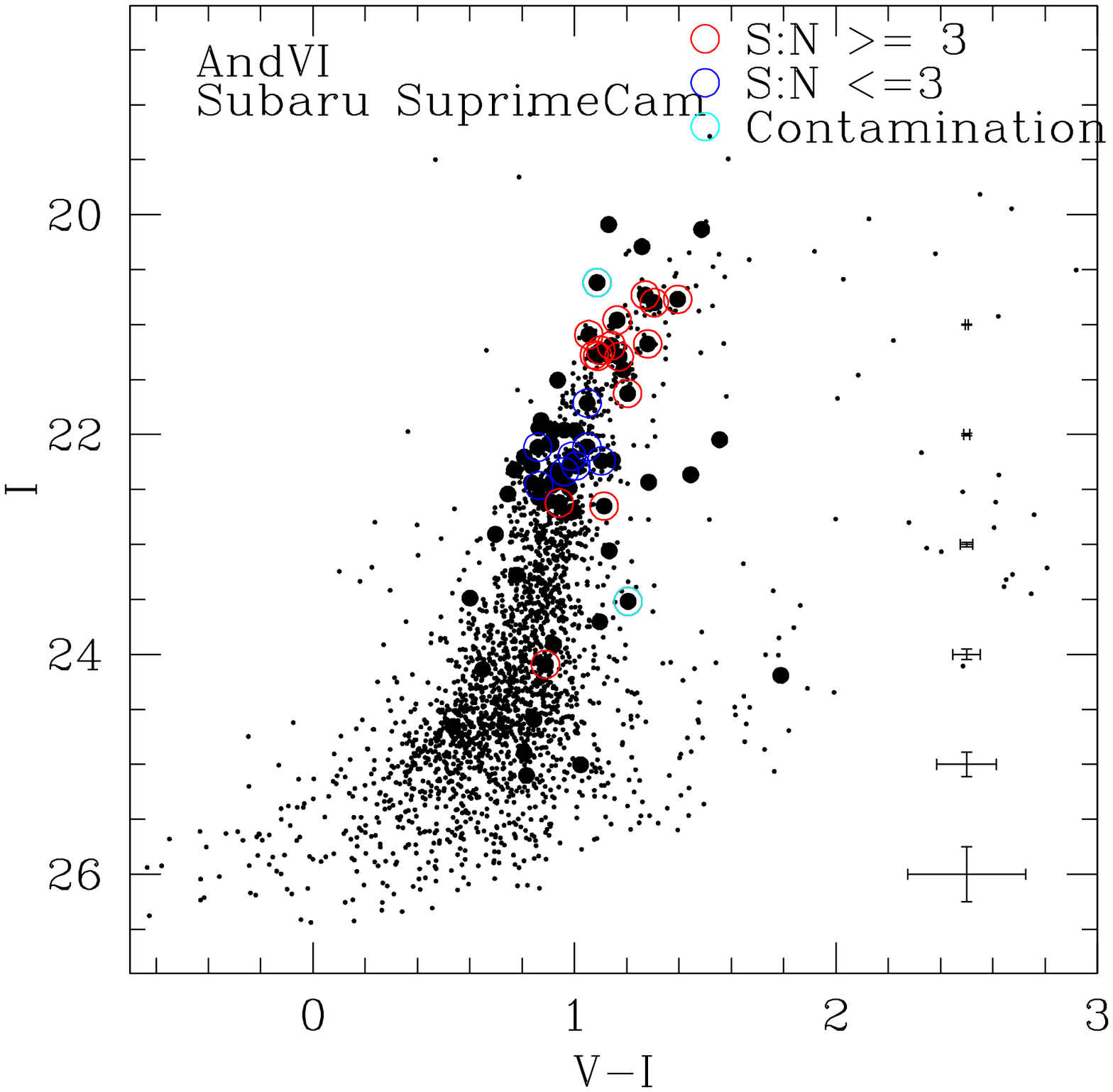}
\caption{Extinction corrected CMDs of And V and And VI (Subaru, $V$
  and $I_c$) for all stellar objects within $1\times r_{half}$ of the
  dSph centres (small black points). In both cases, the RGB is clearly
  seen. Red circles indicate confirmed spectroscopic members for each
  dSph with S:N$>3$, blue circles indicate confirmed spectroscopic
  members for each dSph with S:N$<3$ and cyan circles indicate stars
  with velocities that are consistent with the dSph systemic velocity,
  but are potential M31 halo contaminants. Larger black points
  indicate stars that were observed spectroscopically, but whose
  velocities are inconsistent with the systemic velocities for And V
  and And VI. Error bars show the average 1$\sigma$ uncertainties in
  the photometry at each magnitude level}
\label{cmds}
\end{center}
\end{figure*}

\subsection{And V}

\subsubsection{Structural properties with Subaru Suprime-Cam}

\begin{figure*}
\begin{center}
\includegraphics[angle=0,width=0.45\hsize]{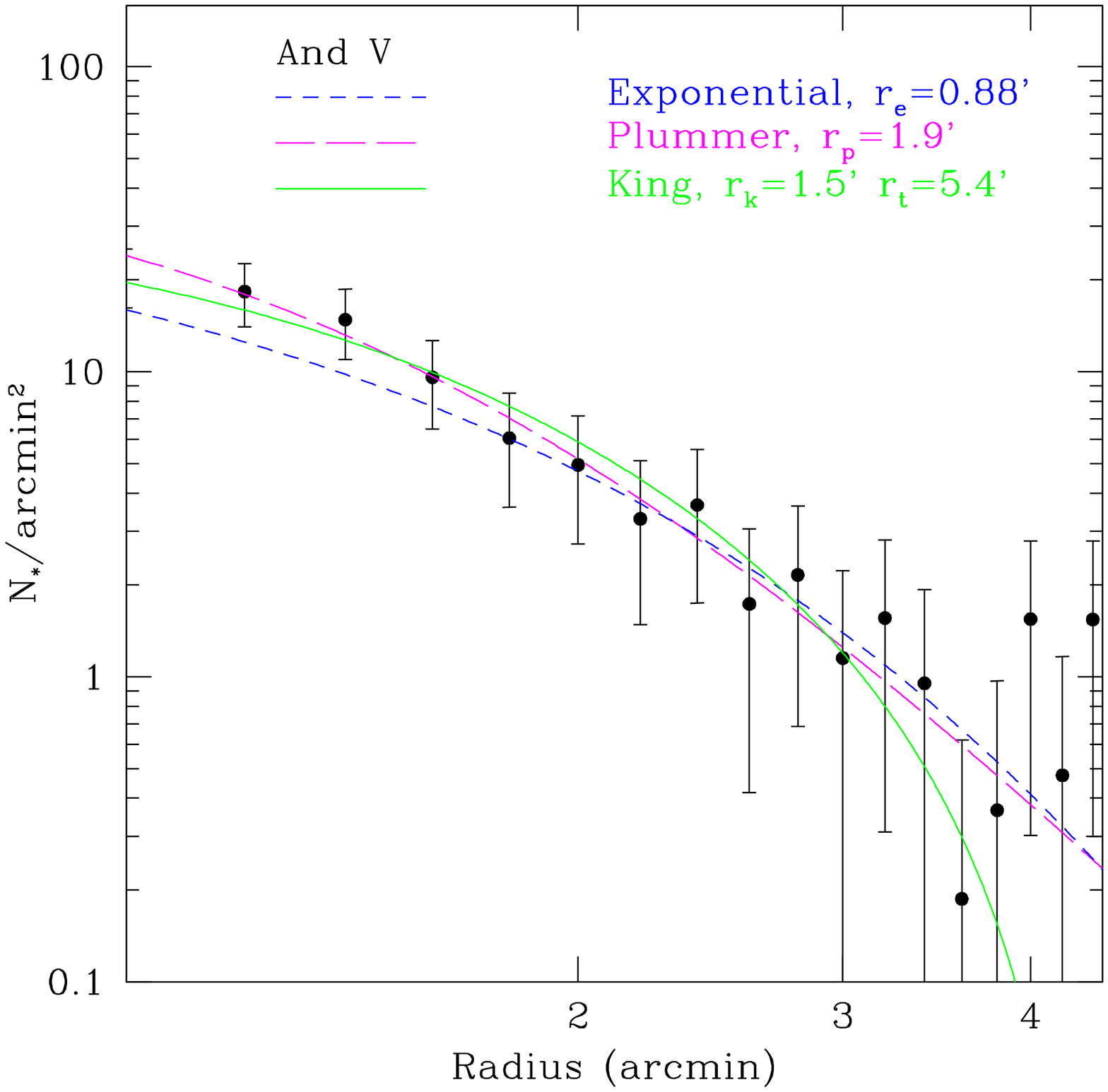}
\includegraphics[angle=0,width=0.45\hsize]{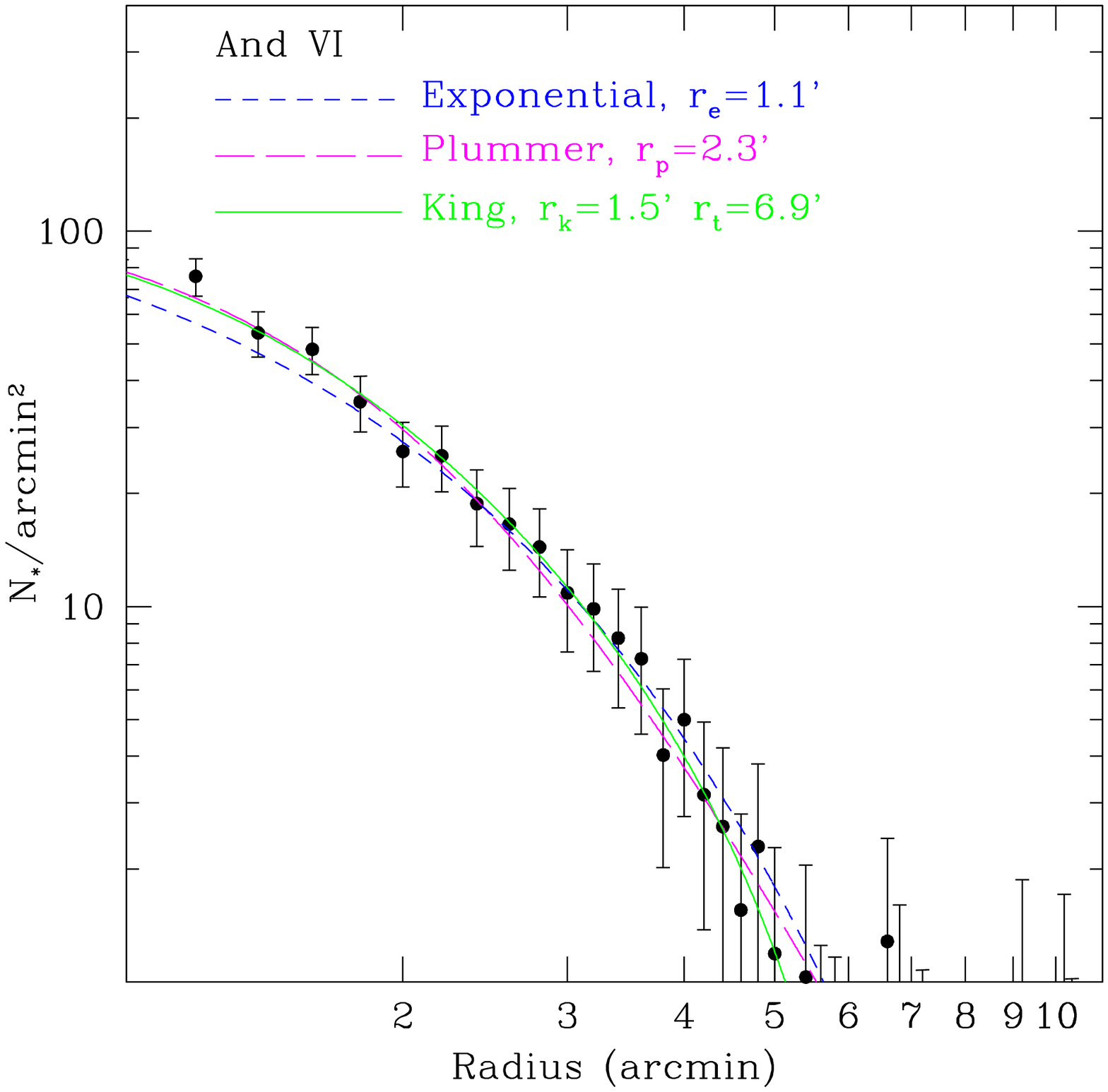}
\caption{ Radial profile for And V (left) and And VI (right)
  constructed from aperture star counts in the region of both dSphs. In
  each case, an average background number density was estimated from
  an annulus located between $8'-12'$ from the centre of each object, and the
  results were subtracted from our star counts. We overlay the best
  fit Exponential (blue short dashed line), Plummer (magenta long-dash
  line) and King (green solid line) profiles in each case also.}
\label{rprofile}
\end{center}
\end{figure*}

Andromeda V is a bright M31 satellite with $M_V=-9.8$
\citep{mcconnachie06b}, and was first discovered with the Second
Palomar Sky Survey \citep{armandroff98}. Its detailed structural
properties were assessed using Isaac Newton Telescope (INT) data by
\citet{mcconnachie06b}. This imaging was performed in Johnson $V$ and
Gunn $i$-band with the Wide Field Camera (WFC), and was deep enough to
observe the top few magnitudes of the red giant branch (RGB) for this
object allowing the measurement of the surface brightness profile,
intensity-weighted centre, position angle, ellipticity and scale-radii
of the dSph using resolved counts of RGB stars. We summarize their
results in Table~\ref{fprops5}. During the nights of August 3--5 2005,
deep Subaru Suprime-Cam imaging of a number of Andromeda dSphs,
including And V were obtained (P.I. N. Arimoto) in the Cousins $V-$
and $I_c-$bands. Conditions were photometric throughout with typical
seeing of $0.5''$. Full details regarding the observing strategy and
reduction techniques for this survey are outlined in
\citet{mcconnachie07}, but we briefly summarize here. Objects were
typically observed in $5\times440$ and $20\times240$ second exposures
in $V$ and $I_c$ bands respectively, allowing equivalent depths to be
reached in both bands. Data were processed using the CASU photometric
pipeline \citet{irwin01}, which de-biased, flat-fielded, trimmed and
gain corrected the images. Then, a catalogue was generated for each
image frame, morphologically classifying each object as stellar,
non-stellar and noise-like. We show the colour magnitude diagram (CMD)
generated from the stellar--classified objects from these
observations, where we have corrected for extinction and reddening
using the maps of \citet{schlegel98}, in Fig.~\ref{cmds}. We now use
this dataset to revise the structural properties of And V. To do this,
we follow the same star count method detailed in
\citet{mcconnachie06a} so that we can make a like-for-like comparison
of the Subaru and INT data. Briefly, we construct an isopleth map of a
$60'\times60'$ field from the Subaru data centred on And V, then
determine the centre of gravity, $\alpha_0,\delta_0$, position angle
$\theta=32\pm2\deg$ (measured from East to North), and ellipticity,
$\epsilon=0.17\pm0.02$, for each isophote by using the
intensity-weighted moments and \citet{mcconnachie06a} equations 2 and
3. We then construct a background corrected radial profile for And V,
where a background level is estimated by measuring the average number
of stars per arcmin within circular annuli located beyond the tidal
radius of And V (as derived in \citealt{mcconnachie06b}). We then
subtract this average background from star counts performed in
elliptical annuli based on the centre of the dSph, and fit the
resulting profile with exponential, Plummer and King profiles. Both
the radial profile and best-fit models are displayed in
Fig.~\ref{rprofile}. From this, we estimate the half-light radius
(using our result for the exponential scale--radius, $r_e=0.88'$) to
be $r_{half}=1.3'$ and (for the King model) tidal radius of $r_t=5.4'$
Using the distance modulus of 24.44 (774 kpc,\citealt{mcconnachie05a})
this gives $r_{half}=292\pm22$~pc and $r_t=1.2\pm0.1\kpc$. Our results
are also summarised in Table~\ref{fprops5}.

We inspected the photometric metallicities of the And V stars using
the Dartmouth isochrone models (Dotter et al. 2008).  We select an age
of 10 Gyrs and [$\alpha$/Fe]=+0.2 as numerous studies of dwarf
spheroidals have shown them to be composed of old stellar populations
that are enhanced in $\alpha$-elements.  We interpolate between these
isochrones on a fine grid, and present the resulting metallicity
distribution function (MDF) in Fig.~\ref{mdf}. We find a mean
metallicity of [Fe/H]=-1.6 for the And V members and a dispersion of
0.3 dex. However, as our imaging is not deep enough to realise the
main sequence turn-off (MSTO) of And V, we cannot reliably ascertain
the precise ages and $\alpha$-abundances of the stellar populations.
The effect of increasing (decreasing) the age used for our isochrones
by $\sim2$Gyr results in a shift of $-0.1$ ($+0.1$) dex to our mean
metallicities.  Similarly, and increase (decrease) in our assumed
value of [$\alpha$/Fe] by 0.2 dex shifts our mean metallicity by
$-0.1$ (+0.1) dex. This results in a combined uncertainty of $\pm0.2$
dex. 

\subsubsection{Kinematic properties with Keck I LRIS}

And V was observed using the Low Resolution Imaging Spectrograph
(LRIS), situated on the Cassegrain focus of the Keck I telescope on
Mauna Kea, on 16th August 2009. The 831/8200 grating was employed with
the $I$-band filter on the LRIS Red II detector system, giving a
resolution of $\sim3.0$\AA\ (R$\sim2800$). Our observations were taken
in an average of 0.7$''$ seeing, covering a wavelength range of
6900--9000\AA, in the region of the Calcium Triplet (Ca II).  We took
4$\times$15 minute exposure, which resulted in typical signal-to-noise
ratios of S:N=3--10\AA$^{-1}$.  Targets were selected using the Subaru CMD
shown in Fig.~\ref{cmds}, prioritising stars with $20.3<i<23.0$ from
within the locus of the clearly defined the RGB.  In total, we
observed 45 science targets within the And V mask, 38 of which reduce
successfully.

To reduce the LRIS data, we used the IRAF noao.twodspec and
noao.onedspec packages.  The spectra were flat-fielded, de-biased and
wavelength calibrated, as well as cleaned of skylines and cosmic
rays. We then performed heliocentric velocity corrections to each of
our observed stars. Our velocities and errors were derived from the Ca
II triplet lines located at $\sim8500$\AA, using the same technique
described in \citet{collins10}.  Briefly, we used an error-weighted
cross-correlation technique with a model template of the Ca II feature at
the rest-frame wavelength positions of the triplet lines. We repeated
the cross-correlation 1000 times, adding random Poisson noise to our
spectra on each occasion, and took the average and standard deviations
of these as our velocities and associated errors.  We then combine
these errors with ones derived by performing separate
cross-correlations to each of the Ca II lines in turn, using the
dispersion of the resulting velocities. This results in typical errors
of $5-15\kms$, with a mean for our sample of 6.8$\kms$.

\subsection{ And VI}

\subsubsection{Structural properties with Subaru Suprime-Cam}

And VI is one of the more luminous of the M31 classical dSphs, with
$M_V=-11.5$ \citep{mcconnachie06b}, and it too was discovered using
the Second Palomar Sky Survey \citep{armandroff99}. Its structural
properties were assessed from INT WFC photometry by
\citet{mcconnachie06a} in the same manner as And V.  These results are
summarised in Table~\ref{fprops}. And VI was also observed in our
Subaru Suprime--Cam survey, and the data were taken and reduced in the
same way as described above. We display the CMD of stellar objects
from this data within 1$\times r_{half}$ of And VI in
Fig.~\ref{cmds}. The data start to become incomplete at a S:N of 10,
which corresponds to $V=I\sim25.5$.

In this section, we re-derive the structural parameters for And VI,
using this Subaru data and the same approach as detailed for And
V. Again, from the star counts and isopleths maps we find a good
agreement with those of the INT study. Our results are summarised
in Table~\ref{fprops}. We calculate $r_{half}=1.9'$ (from our
exponential scale radius), $r_t=7.0'$, $\epsilon=0.39$ and
$\theta=164\deg$. Using the \citet{mcconnachie05a} distance for And VI
of 783 kpc (distance modulus of 24.47) we obtain physical scale radii
of $r_{half}=440\pm16$~pc and $r_t=1.6\pm0.2\kpc$.

\begin{figure*}
\begin{center}
\includegraphics[angle=0,width=0.45\hsize]{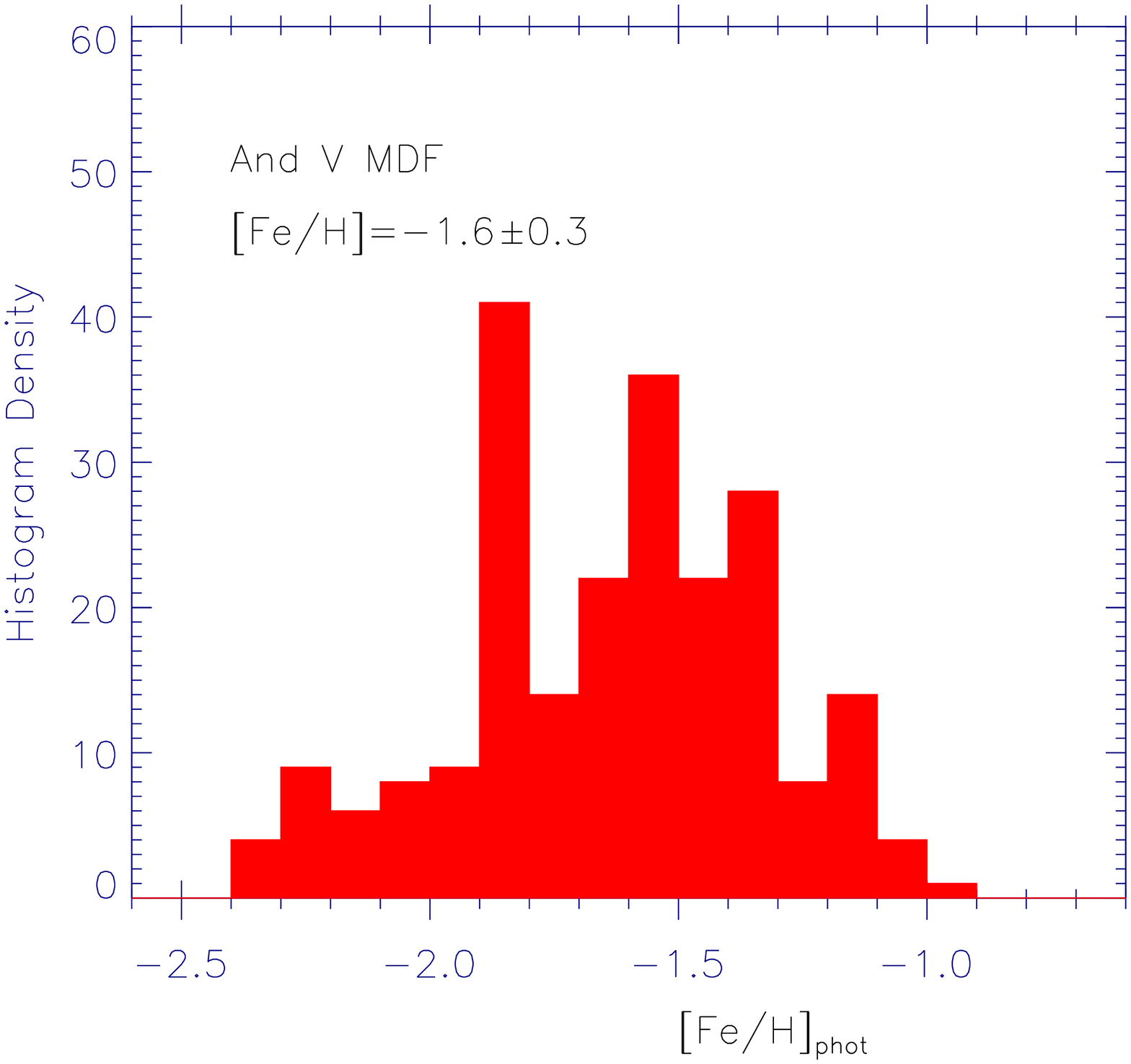}
\includegraphics[angle=0,width=0.45\hsize]{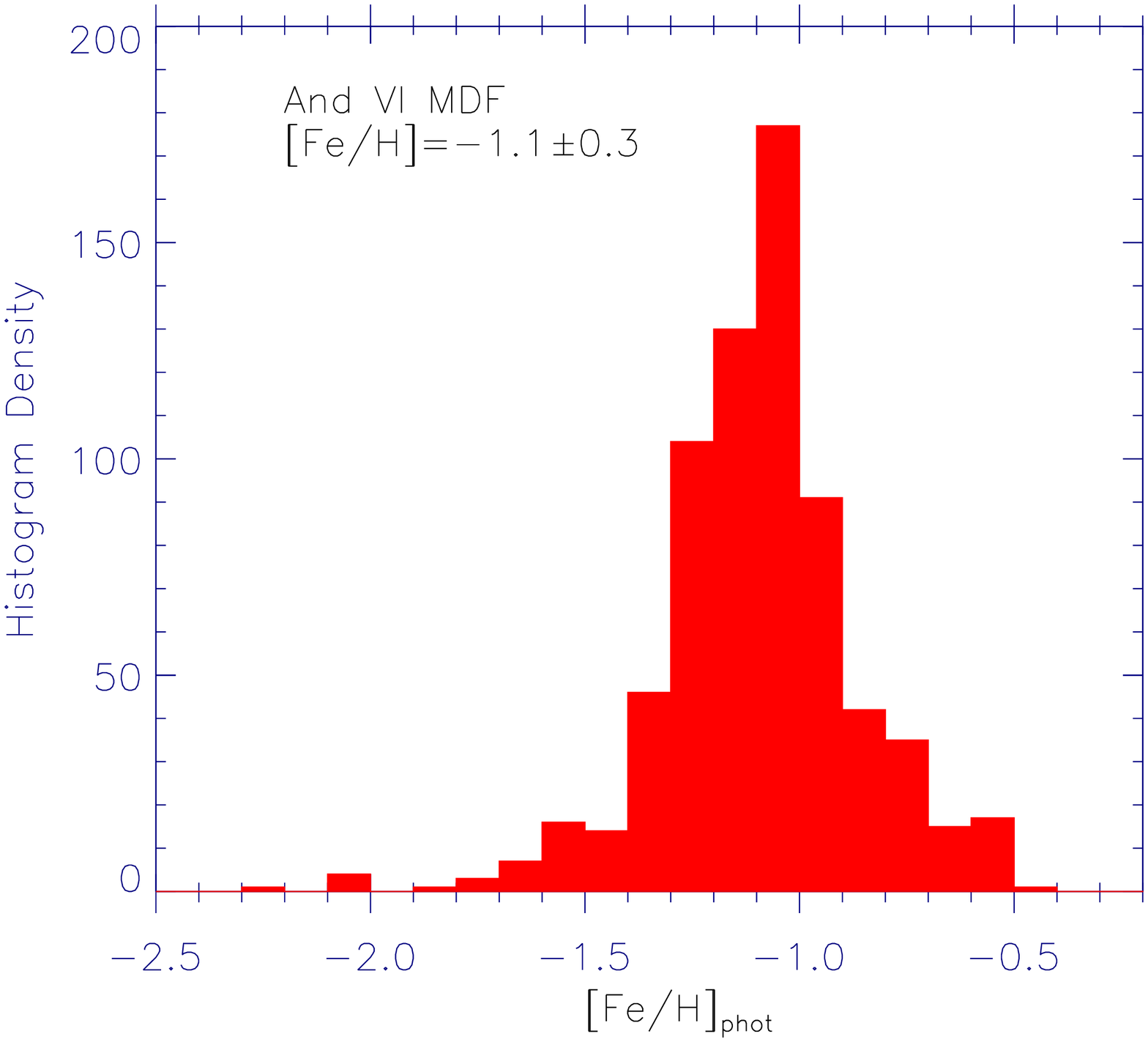}
\caption{The MDF for And V and And VI, constructed from metallicities
  for stars within 2$r_{half}$ of the centre of each dSph. These were
  calculated by interpolating between a fine grid of Dartmouth
  isochrones \citep{dart08} with [$\alpha$/Fe]=+0.2 and an age of 10
  Gyr. The mean metallicity is found to be $\feh=-1.6\pm0.3$ and
  $\feh=-1.1\pm0.3$ dex for And V and And VI respectively where the
  error represents the dispersion of the population. }
\label{mdf}
\end{center}
\end{figure*}

We perform the same isochrone analysis for And VI as that described in
\S2.1.1, and the resulting MDF is displayed in Fig.~\ref{mdf}. The
same difficulties in ascertaining robust photometric metallicities for
And V are experienced here, as the Subaru data do not reach the
MSTO. We find a median metallicity of [Fe/H]$=-1.1\pm0.3$, where
the error represents the dispersion of the population.  This
dispersion is the same as we measure for And V, despite And VI having
a much broader RGB. And V is more metal poor than And VI with
an average metallicity of [Fe/H]$=-1.6$. As you decrease in
metallicity, isochrones for a given age and $\alpha$-abundance begin
to bunch together in colour space ($V-I$), meaning smaller differences
in colour translate to larger differential metallicities, driving a
larger metallicity dispersion for a smaller colour range.

\subsubsection{Kinematic properties with Keck II DEIMOS}

The DEep-Imaging Multi-Object Spectrograph (DEIMOS), situated on the
Nasmyth focus of the Keck II telescope is an ideal instrument for
obtaining medium resolution (R$\sim6000$) spectra of multiple, faint
stellar targets in the M31 dSphs. The data for And VI were taken on
the nights of 17 --19 September 2009 in photometric conditions and
with $<1''$ seeing.  Our chosen instrumental setting covered a
wavelength range of 5600--9800\AA\, and for our exposures we
implemented 3$\times$30 minute integrations, and employed the 1200
line/mm grating, giving us a spectral resolution of $\sim1.37$\AA. The
spectra from this setup typically possess S:N of $>3$\AA$^{-1}$. Due
to the position of And VI in the sky, this object was observed at the
beginning of each night meaning that it suffered from a high airmass
of 2.2, so the S:N for these stars are lower than expected for
comparable length exposures taken at a lower airmass. The typical
errors within the mask range from $5-12\kms$, with a mean for our
sample of 6.2$\kms$.

\subsection{Determining membership}

Before we can analyse the kinematic properties of these objects, it is
important to correctly determine the bona-fide members of the dSphs,
and to eliminate stars which belong to the MW foreground (mostly $v_r
>-160\kms$) or the M31 halo ($v_r\sim-300\kms$). As both And V and
And VI lie at a large projected radius from the centre of M31
($\sim110$ and 270 kpc respectively,
\citealt{armandroff98,armandroff99}), the density of stars, and thus
the number of contaminants, belonging to the M31 halo should be very
low for these objects, but should still be considered. As And V is
located at a higher latitude than And VI, this dwarf is more likely to
experience contamination from the MW, and so this must be treated
carefully.In order to minimise the contamination to our sample, we
require the following conditions to be met by bona-fide members:

\noindent$\bullet$ Stars must fall on the RGB of dSph, as defined by
colour cuts in $V$ and $I$- bands for And V and 
And VI.

\noindent$\bullet$ Stars must sit within 2$r_{half}$ of the dSph
centre. Stars at larger radii that satisfy all other criteria barring
this are classified as `tentative members', and their properties and
probabilities of being members will be discussed.

\noindent$\bullet$ Stars must show low Na I doublet absorptions
(i.e. EW$_{NaI}<1.4$). Strong Na I absorption may indicate that the
observed star is a foreground dwarf star, not an M31 RGB star.

\noindent$\bullet$ After deriving a systemic velocity, any star within
the sample with a velocity that is $>3\sigma$ from the systemic will
not be considered a member. The kinematic properties of the dSph will
then be re-derived iteratively in this manner until only stars
with velocities within $3\sigma$ of the derived systemic velocity remain.

\begin{table}
\begin{center}
\caption{Structural properties of And V derived in McConnachie et al. 2006a and this work.}
\label{fprops5}
\begin{tabular}{lcc}
\hline
Property & M06a & This work  \\
\hline
$\alpha_0$ (h:m:s)         & 01:10:17.0     & 01:10:17.9     \\
$\delta_0$ ($\deg:':''$)   &+47:37:46       & +47:37:38.0    \\
$\theta$   ($\deg$)        & 32$\pm3$       & 32$\pm2$       \\
$\epsilon$                 & 0.18$\pm0.03$  & 0.17$\pm0.02$  \\
$r_e$      (arcmin)        & 0.86$\pm0.05$  & 0.88$\pm0.09$  \\
$r_p$      (arcmin)        & 1.56$\pm0.08$  & 1.59$\pm0.1$  \\
$r_c$      (arcmin)        & 1.2$\pm0.2$    & 1.2$\pm0.1$    \\
$r_t$      (arcmin)        & 5.3$\pm0.4$    & 5.4$\pm0.8$    \\
$r_t$      (kpc)           & 1.2$\pm0.2$    & 1.2$\pm0.2$               \\
$r_{half}$      (arcmin)    & 1.3            & 1.3$\pm0.1$\\
$r_{half}$      (pc)        & 300            & 292$\pm22$ \\
$\feh_{phot}^{(a)}$          &  -1.6           & -1.6$\pm0.3$   \\  
\hline
\end{tabular}
\end{center}
$^{(a)}$ Error represents dispersion.
\end{table}

\begin{table}
\begin{center}
\caption{Structural properties of And VI derived in McConnachie et al. 2006a and this work.}
\begin{tabular}{lcc}
\hline
Property & M06a & This work \\
\hline
$\alpha_0$ (h:m:s)         & 23:51:46.9      & 23:51:47.3    \\
$\delta_0$ ($\deg:':''$)& +24:34:57       & +24:34:52.5      \\
$\theta$   ($\deg$)     & 163$\pm3$       & 164$\pm2$        \\
$\eta$                     & 0.41$\pm0.03$   & 0.39$\pm0.02$ \\
$r_e$      (arcmin)        & 1.2$\pm0.04$    & 1.1$\pm0.04$  \\
$r_p$      (arcmin)        & 2.15$\pm0.08$   & 2.3$\pm0.08$  \\
$r_c$      (arcmin)        & 2.1$\pm0.2$     & 1.5$\pm0.1$   \\
$r_t$      (arcmin)        & 6.2$\pm0.4$     & 7.0$\pm0.8$   \\
$r_t$      (kpc)           & 1.4$\pm0.1$     & 1.6$\pm0.2$   \\
$r_{half}$      (arcmin)        & 1.8         & 1.9$\pm0.07$  \\
$r_{half}$      (pc)           & 420        & 440$\pm16$ \\
$\feh_{phot}^{(a)}$         &  -1.3           & -1.1$\pm0.3$   \\  
\hline
\label{fprops}
\end{tabular}
\end{center}
$^{(a)}$ Error represents dispersion.
\end{table}

\section{The kinematics of \andv\ and VI}

\subsection{Systemic velocities and velocity dispersions}

\begin{figure*}
\begin{center}
\includegraphics[angle=0,width=0.4\hsize]{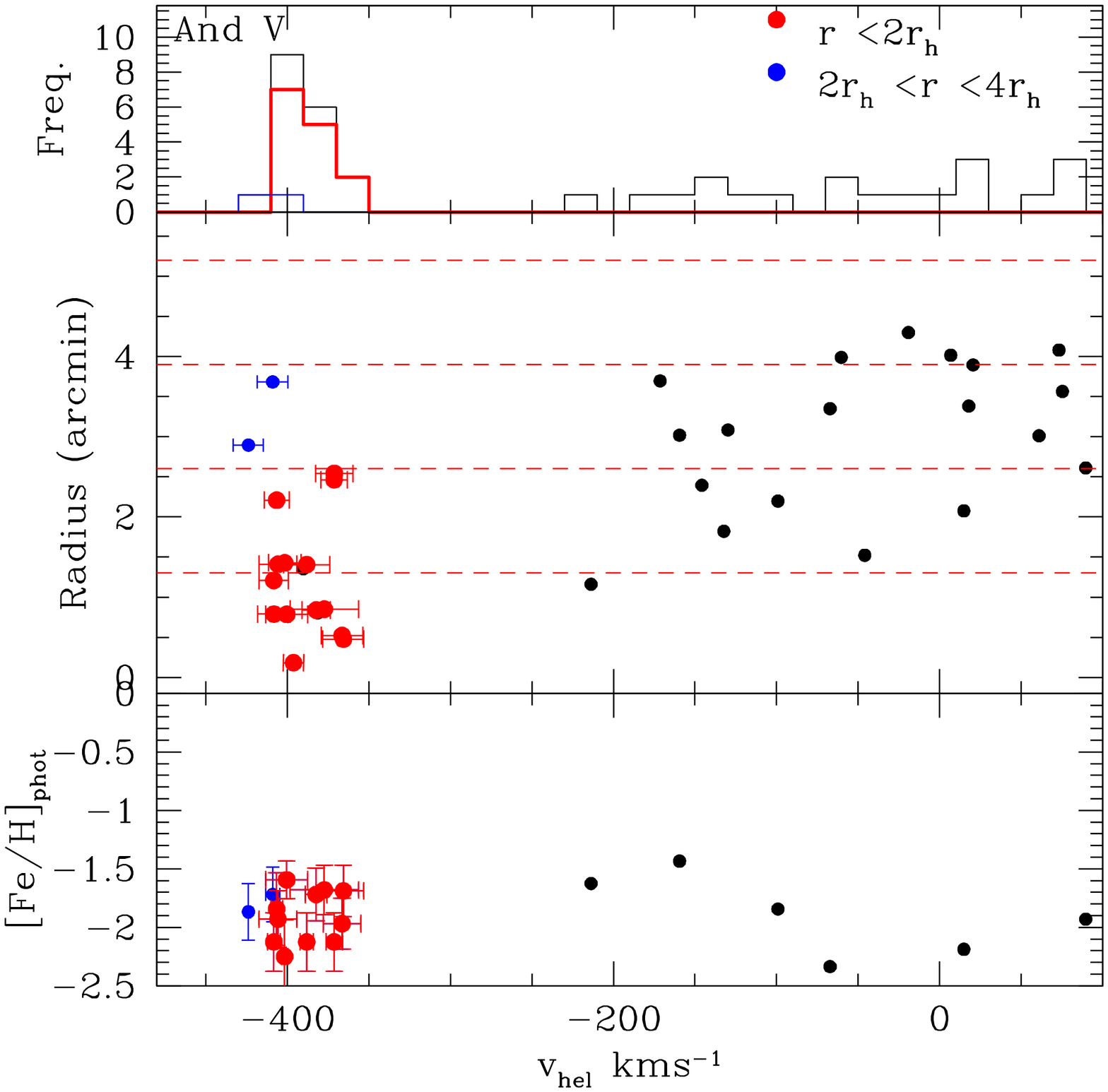}
\includegraphics[angle=0,width=0.4\hsize]{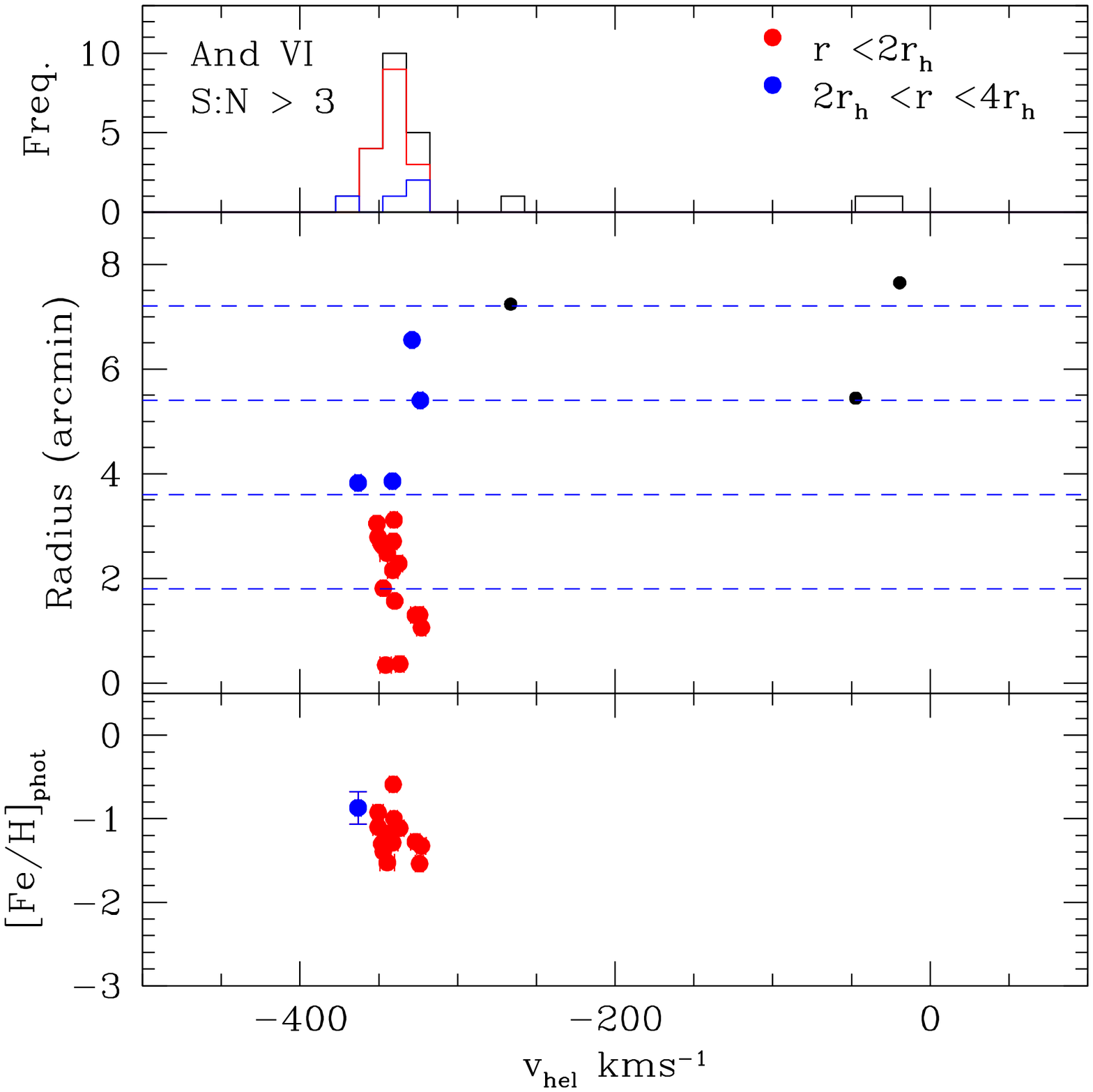}
\caption{The velocities, distances from centre and
  photometric metallicities of all stars within the LRIS and DEIMOS
  fields. Some stars do not have photometric metallicities as their
  colours place them outside the range of isochrones used. Red points
  indicate likely dSph members and the red dashed lines represent 1,
  2, 3 and 4 times the r$_h$ of each dSph. And V and And VI appear as
  cold over--densities of stars at $\sim-400\kms$ and $\sim-350\kms$
  respectively.}
\label{vels}
\end{center}
\end{figure*}

\begin{table*}
\begin{center}
\caption{Kinematic properties of observed members of And V}
\label{andv}
Confirmed members\\
\begin{tabular}{cccccccc}
\hline
$\alpha_{2000} $(hh:mm:ss)& $\delta_{2000}$ & $V$ & $I$ &  $v_r (\kms$) & S:N (\AA$^{-1}$) & [Fe/H]$_{phot}^{(a)}$& $\feh_{spec}$\\
\hline
1:10:18.62 & 	47$\deg$38$'$9.4$''$ &	22.544 &21.405 &  -365.6$\pm$12.6  &   6.5  &   -1.65  & -2.1$\pm0.3$ \\
1:10:12.19  &	47$\deg$37$'$30.6$''$ &	21.95 &	20.581 &  -377.3$\pm$21    &   5.1  &   -1.55 &  -1.3$\pm$0.3 \\
1:10:20.65  &	47$\deg$38$'$15.7$''$ &	21.979 &20.735 &  -408.2$\pm$4     &   4.6  &   -1.85 &  -2.2$\pm0.4$ \\
1:10:20.03 & 	47$\deg$37$'$10.1$''$ &	22.349 &21.101 &  -400.4$\pm$12.8  &   4.4  &   -1.48 &  -1.2$\pm0.3$ \\
1:10:16.32  &	47$\deg$37$'$37.4$''$ &	22.803 &21.815 &  -396.2$\pm$2.21  &   4.3  &   -2.15 &  -1.6$\pm0.7$ \\
1:10:22.22  &	47$\deg$38$'$52.1$''$ &	22.685 &21.571 &  -405.7$\pm$11.5  &   4.2  &   -1.55 &  -2.1$\pm0.3$ \\
1:10:25.14  &	47$\deg$38$'$9.8$''$ &	22.875 &21.873 &  -401.6$\pm$1.94  &   3.9  &   -2.05 &  -2.0$\pm0.3$ \\
1:10:17.94  &	47$\deg$37$'$16.2$''$ &	22.05 &	20.804 &  -366.4$\pm$11.7  &   3.5  &   -1.78 &  -1.0$\pm0.4$ \\
1:10:06.71  &	47$\deg$35$'$26.9$''$ &	22.218 &20.986 &  -423.8$\pm$1.28  &  3.4  &    -1.65 &  -1.8$\pm0.4$ \\
1:10:09.04  &	47$\deg$37$'$20.8$''$ &	22.294 &21.125 &  -388.1$\pm$4.1   &   3.4  &   -1.78 &  -1.6$\pm0.5$ \\
1:10:24.17  &	47$\deg$37$'$47.1$''$ &	21.323 &19.909 &  -408.3$\pm$9.04  &   2.8  &   -2.12 &  -1.7$\pm0.7$ \\
1:10:21.56  &	47$\deg$37$'$25.8$''$ &	22.152 &20.86 &	  -382.3$\pm$6.76  &   2.2  &   -1.55 &  -1.8$\pm0.5$ \\
1:10:05.54 & 	47$\deg$36$'$41.6$''$ &	22.178 &20.871 &  -406.5$\pm$3.6   &   2.0  &   -1.48 &  -1.4$\pm0.5$ \\
1:10:02.38 &	47$\deg$37$'$48.5$''$ &	22.84 &	21.808 &  -371.3$\pm$5.08  &   1.7  &   -1.85 &  -2.2$\pm0.5$ \\
\hline
\end{tabular}
\\Tentative members (i.e. stars at $>2\times r_{half}$)
\begin{tabular}{cccccccc}
$\alpha_{2000} $(hh:mm:ss)& $\delta_{2000}$ & $V$ & $I$ &  $v_r (\kms$) & S:N (\AA$^{-1}$) & [Fe/H]$_{phot}^{(a)}$& $\feh_{spec}$\\
\hline
1:10:07.45 & 	47$\deg$35$'$48.0$''$ & 21.845 &20.526 &  -371.2$\pm$2.34  &  3.9  &    -1.9 &   -1.9$\pm0.4$ \\
1:10:38.56  &	47$\deg$38$'$22.1$''$ &	22.51 &	21.346 &  -409.0$\pm$4.38  &   3.0  &   -1.55 &  -2.0$\pm0.4$ \\
\hline
\end{tabular}
\\Outliers\\
\begin{tabular}{cccccccc}
$\alpha_{2000} $(hh:mm:ss)& $\delta_{2000}$ & $V$ & $I$ &  $v_r (\kms$) & S:N (\AA$^{-1}$) & [Fe/H]$_{phot}^{(a)}$& $\feh_{spec}$\\
\hline
1:10:19.23  &	47$\deg$39$'$4.1$''$ &	21.748 &20.563 &  -390.2$\pm$1.32  &   3.8  &   -2.42 &    -1.3$\pm0.5$ \\
1:10:13.94  &	47$\deg$37$'$9.0 $''$ & 22.067 &20.989 &  -381.3$\pm$2.03  &   3.7  &   -2.48  &   -1.6$\pm0.7$ \\
\hline
\end{tabular}
\end{center}
$^{(a)}$ Derived from \citet{dart08} isochrones with [$\alpha$/Fe]=+0.2, age=10 Gyrs. Typical errors of $\pm0.2$ dex.
\end{table*}

\begin{table*}
\begin{center}
\caption{Kinematic properties of observed members of And VI}
\label{andvi}
Confirmed members\\
\begin{tabular}{cccccccc}
\hline
$\alpha_{2000} $(hh:mm:ss)& $\delta_{2000}$ & $V$ & $I$ &  $v_r (\kms$) & S:N (\AA$^{-1}$) & [Fe/H]$_{phot}^{(a)}$& $\feh_{spec}$\\
\hline
   23:51:44.45  &   24$\deg$37$'$48.80$''$ &    22.12   &  20.96 &  -347.25$\pm$1.73  & 5.02 & -1.22 &  -2.04 $\pm$0.8\\ 
   23:51:50.74  &   24$\deg$33 $'$57.70$''$ &    23.76 &    22.65 &  -340.84$\pm$2.36 & 4.57 & -0.70 & -0.96$\pm$0.5\\ 
   23:51:45.55  &   24$\deg$36$'$ 3.80$''$ &    22.15   &  21.09  & -324.21$\pm$2.57  & 4.38 & -1.35 & -0.71$\pm$0.2\\ 
   23:51:52.41  &   24$\deg$33$'$34.70$''$ &   22.83  &   21.63  & -340.42$\pm$2.14   & 4.04 & -0.80 & -1.41$\pm$0.5\\ 
   23:51:51.52  &   24$\deg$36$'$23.50$''$ &    22.46  &   21.29  & -345.70$\pm$3.48  & 3.84 & -1.04 & -0.98$\pm$0.2\\ 
   23:51:46.94   &  24$\deg$35$'$40.70$''$ &    22.33  &   21.19  & -326.74$\pm$3.01  &3.81  & -1.12 & -1.71$\pm$0.4\\ 
   23:51:48.19  &   24$\deg$34$'$27.10$''$ &   24.98   &  24.09  & -337.44$\pm$2.64   & 3.57 & -0.92 & -1.15$\pm$0.4\\ 
   23:51:41.15   &  24$\deg$36$'$33.00$''$ &   22.38   &  21.29  & -341.16$\pm$3.32   & 3.57 & -1.15 & -0.78$\pm$0.3\\ 
   23:51:46.60  &   24$\deg$37$'$11.90$''$ &    22.35  &   21.27  & -323.01$\pm$2.88  & 3.56 & -1.18 & -2.45$\pm$2.2\\ 
   23:51:50.85  &   24$\deg$36$'$18.50$''$ &    22.46   &  21.18 &  -336.70$\pm$2.58  & 3.55 & -0.99 & -1.81 $\pm$0.6\\ 
   23:51:44.38  &   24$\deg$34$'$16.10$''$ &    24.06  &   23.28 &  -350.49$\pm$3.40  & 3.20 & -0.73 & -0.78 $\pm$0.4\\ 
   23:51:46.21  &   24$\deg$37$'$37.10$''$ &   23.61    & 22.91  & -347.09$\pm$5.88   & 3.10 & -1.18 & -1.37$\pm$0.3\\ 
   23:51:46.19  &   24$\deg$35$'$54.60$''$ &   25.70   &  24.90  &  -332.98$\pm$6.07  & 2.08 & -1.15 & -1.09$\pm$0.4\\ 
   23:51:41.12  &   24$\deg$37$'$43.50$''$ &    23.00   &  22.09  & -308.26$\pm$8.37  & 2.00    & -1.05 & -1.59 $\pm$3.2\\ 
   23:51:54.77  &  24$\deg$28$'$53.50$''$ &    23.19   &  22.20  & -339.81$\pm$6.66   & 1.90 & -0.80 & -1.87$\pm$0.9 \\
   23:51:46.91  &   24$\deg$33$'$53.00$''$ &    23.69  &   22.70 &  -314.51$\pm6.52$  & 1.89  & -0.95 & -0.95$\pm$0.7\\ 
   23:51:54.79  &   24$\deg$35$'$50.30$''$ &    22.88  &   21.96 &  -328.10$\pm$7.15  & 1.86 & -1.10 & -1.41 $\pm$1.1\\ 
   23:51:44.42  &   24$\deg$37$'$26.30$''$ &    23.38   &  22.46  & -348.25$\pm$6.95  & 1.46 & -0.83 & 1.57$\pm$ 11.1\\ 
   23:51:54.23   &  24$\deg$36$'$42.80$''$ &    22.80  &   21.94  & -351.77$\pm$12.16 & 1.43 & -1.25 & -1.7$\pm$0.9\\ 
   23:51:44.13  &   24$\deg$35$'$45.10$''$ &    23.37  &   22.44  & -332.82$\pm$8.19  & 1.39 & -0.80 & -1.14$\pm$0.6\\ 
   23:51:49.63  &   24$\deg$32$'$42.20$''$ &    22.98  &   22.12  & -354.98$\pm$8.92  & 1.35 & -1.15 & -2.16$\pm$11.4\\ 
   23:51:44.66   &  24$\deg$36$'$37.90$''$ &    24.83   &  23.91 &  -340.10$\pm$35.83 & 1.21 & -1.15 & -0.54$\pm$0.5\\ 
   23:51:46.70   &  24$\deg$35$'$16.50$''$ &    24.09  &   23.49  & -354.44$\pm$11.69 & 1.33 & -1.25 &  -1.88$\pm$0.7\\ 
   23:51:45.61  &   24$\deg$36$'$47.10$''$ &    22.92   &  21.96 &  -362.91$\pm$8.38  & 1.11 & -1.02 & -1.16$\pm$0.9\\ 
   23:51:44.69  &   24$\deg$34$'$42.30$''$ &    25.44   &  24.59 &  -384.50$\pm$9.38  & 0.89 & -0.90 & -0.87$\pm$0.9\\ 
   23:51:41.55   &  24$\deg$36$'$28.10$''$ &   23.27    & 22.44  & -320.62$\pm$14.57  & 0.87 & -1.05 & -1.03$\pm$0.9 \\
\hline
\end{tabular}
\\Tentative members (i.e. stars at $>2\times r_{half}$)
\begin{tabular}{cccccccc}
$\alpha_{2000} $(hh:mm:ss)& $\delta_{2000}$ & $V$ & $I$ &  $v_r (\kms$) & S:N (\AA$^{-1}$) & [Fe/H]$_{phot}^{(a)}$& $\feh_{spec}$\\
\hline
   23:51:47.00   &  24$\deg$32$'$56.00$''$ &    23.56  &   22.62  & -363.18$\pm$1.85  & 4.71 & -0.68 & -1.20$\pm$0.3\\ 
   23:51:45.80  &   24$\deg$32$'$31.20$''$ &    22.76  &   21.71  & -326.89$\pm$5.04  & 3.29 & -1    & -1.41$\pm$ 0.4\\ 
   23:51:54.12 &    24$\deg$31$'$36.80$''$ &    23.30  &   22.34 &  -348.75$\pm$5.65  & 3.75 & -0.78 & -1.36$\pm$0.5\\ 
   23:51:49.59  &   24$\deg$31$'$27.20$''$ &    23.38  &   22.23 &  -345.96$\pm$8.11  & 3.51 & -0.57 & -1.10$\pm$0.7\\ 
   23:51:49.30  &   24$\deg$32$'$26.80$''$ &    23.16  &   22.11 &  -343.06$\pm$5.88  & 1.46 & -0.74 &  -2.06$\pm$0.7\\ 
   23:51:45.30  &   24$\deg$30$'$18.50$''$ &   23.34  &   22.24  & -351.28$\pm$ 8.70  & 1.15 &-0.60  & -1.34$\pm$0.8\\
   23:51:47.81 &    24$\deg$31$'$18.10$''$ &    23.29  &   22.29 &  -320.05$\pm$11.83 & 1.10 & -0.73 & -1.97$\pm$0.5 \\
\hline
\end{tabular}
\\Outliers\\
\begin{tabular}{cccccccc}
$\alpha_{2000} $(hh:mm:ss)& $\delta_{2000}$ & $V$ & $I$ &  $v_r (\kms$) & S:N (\AA$^{-1}$) & [Fe/H]$_{phot}^{(a)}$& $\feh_{spec}$\\
\hline
   23:51:44.87  &   24$\deg$37$'$31.90$''$ &    21.70   &  20.62 &  -339.94$\pm$1.55  & 6.39 & NA    & -1.65$\pm$0.3\\ 
   23:51:44.69  &   24$\deg$34$'$42.30$''$ &    24.19   &  23.06 &  -384.50$\pm$9.38  & 0.89 & -0.20 & -1.20$\pm$1.3\\ 
\hline
\end{tabular}
\end{center}
$^{(a)}$ Derived from \citet{dart08} isochrones with [$\alpha$/Fe]=+0.2, age=10 Gyrs. Typical errors of $\pm0.2$ dex.
\end{table*}

Using the criterion laid out in \S~2.3 we identify the most-probable
member stars for each of the dSphs observed with LRIS and DEIMOS. In
Figure~\ref{vels} we display the kinematic properties of all the stars
observed in each slit mask. The top panel in each case shows a
histogram of the velocities of observed stars. The heavy red histogram
highlights the kinematic location of each dSph, which are immediately
obvious as cold over-densities of stars in velocity space. The central
panel shows velocity vs. distance from the dwarf centre, where the 1,
2, 3 and 4$\times~r_{half}$ are marked as dashed lines. The lower
panel shows the velocity as a function of photometric
metallicity. {\bf Not all stars within the mask have photometric
  metallicities as their colours place them outside the parameter
  space covered by our chosen isochrones. Such drop outs are likely
  Galactic contaminants and not true M31 RGB stars}. Cutting on
velocity, distance, NaI absorption and metallicity (proxy for position
on RGB), we identify 14 and 26 secure members per dSph for And V and
And VI respectively, as well as 2 and 7 (4 of which have S:N$>3$)
tentative members beyond $2\times r_{half}$ that fall within the cold
peak, although we treat these cautiously as at the systemic velocities
for these objects ($\sim-400\kms$ and $\sim-350\kms$) fall within the
regime of the M31 halo. In the case of And V, two stars found within
the velocity peak do not fall on the RGB of the dwarf (shown as cyan
points in Fig.~\ref{cmds}), seemingly more metal poor than the And V
population, making them likely halo interlopers. For And VI we also
identify 2 outliers from the CMD which again are likely halo
contaminants. In both cases, the outliers show no significant Na I
absorption, strengthening our assumption that these are M31 halo RGB
stars, rather than Galactic halo dwarfs. We show the kinematic
properties of all confirmed and tentative members, as well as halo
contaminants for And V and And VI in Tables~\ref{andv} and \ref{andvi}
respectively.

\begin{figure*}
\begin{center}
\includegraphics[angle=0,width=0.45\hsize]{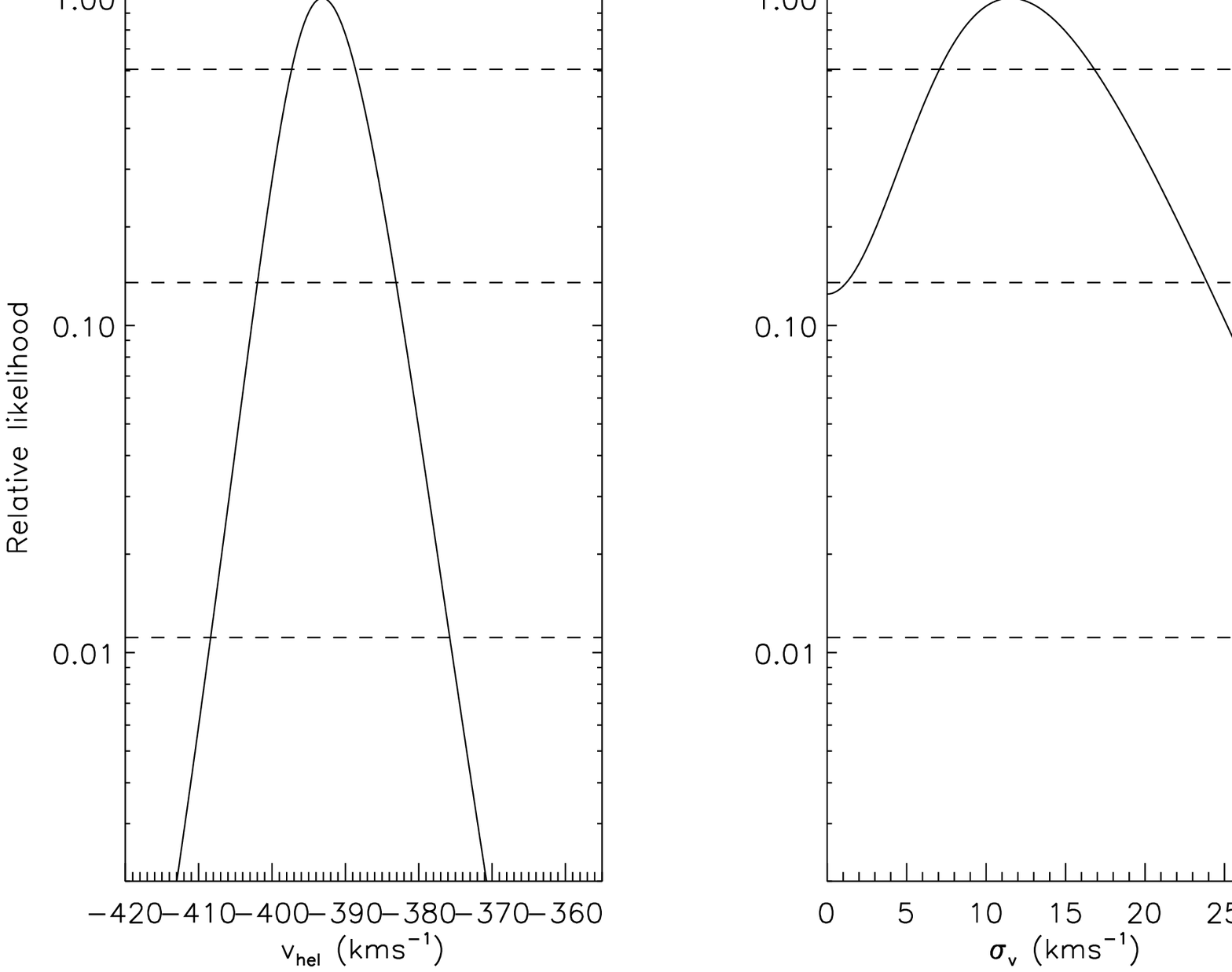}
\includegraphics[angle=0,width=0.45\hsize]{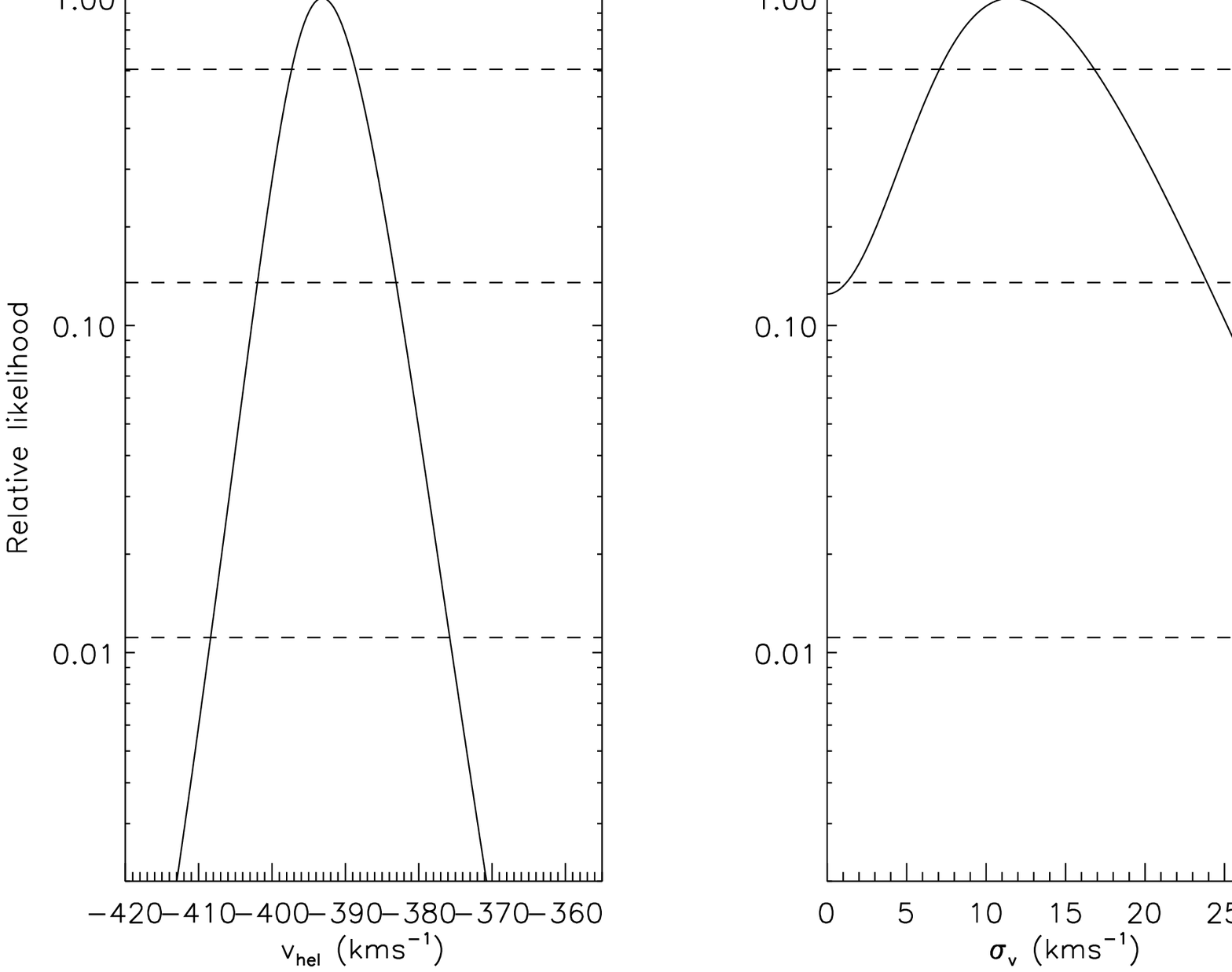}
\caption{ ML distributions for the systemic velocities
  and velocity dispersions of And V (left) and And VI (right). Dashed
  lines represent the 1, 2 and 3$\sigma$ uncertainties on the peak
  values. We measure $v_r=-393.1\pm4.2\kms$ and $-344.8\pm2.5\kms$,
  and$\sigma_v=11.5^{+5.3}_{-4.4}\kms$ and
  $\sigma_v=9.4^{+3.2}_{-2.4}\kms$ for And V and And VI respectively.}
\label{maxlike}
\end{center}
\end{figure*}

With secure candidates in hand, we now determine the systemic
velocities ($v_r$) and velocity dispersions ($\sigma_v$) for each dSph
using the maximum likelihood approach of \citet{martin07}. This is the
same technique as we have used in our previous work
(e.g. \citealt{chapman06,letarte09,collins10}), and we briefly
summarise here. We calculate $v_r$ and $\sigma$ by sampling a coarse
grid in $(v_r,\sigma)$ space and determining the parameter values that
maximise the likelihood function (ML), defined as:

\begin{equation} 
ML(v_r,\sigma)=\prod_{i=1}^{N}\frac{1}{\sigma_{\mathrm{tot}}}
\exp\Big[-\frac{1}{2}\left(\frac{v_r-v_{r,i}}{\sigma_{\mathrm{tot}}}\right)^2\Big]
\end{equation}

\noindent with $N$ the number of stars in the sample, $v_{r,i}$ the
radial velocity measured for the $i^\mathrm{th}$ secure member star,
$v_{err,i}$ the corresponding uncertainty and
$\sigma_{\mathrm{tot}}=\sqrt{\sigma^2+v_{err,i}^2}$. In this way, we
are able to separate the intrinsic dispersion of the dSph from the
dispersion introduced by our measurement uncertainties. We display the
one dimensional likelihood distributions for v$_r$ and $\sigma_v$ in
Fig.~\ref{maxlike} where the dashed lines represent the conventional
1, 2 and 3$\sigma$ (68\%, 95\% and 99.7\%) uncertainties on the
values. In the case of the And VI data, we have only included the 12
member stars with S:N$>3$~\AA\ in this analysis to avoid introducing
large uncertainties into their calculated properties. We determine
systemic velocities of $v_r =-393.1\pm4.2$ and $-344.8\pm3.4\kms$ and
velocity dispersions of $\sigma_v = 11.5^{+5.3}_{-4.4}$ and
$9.4^{+3.2}_{-2.4}\kms$ for And V and VI respectively. The velocities
of these two galaxies (but not dispersions) were previously reported
in \citet{guhathakurta00} to be $v_r=-387.0\pm4.0\kms$ and
$v_r=-340.7\pm2.9\kms$. which agree with our findings within the
associated $1\sigma$ errors. We summarise our results in
Table~\ref{kprops}.

\begin{table}
\begin{center}
\caption{Kinematic properties of And V and And VI.}
\label{kprops}
\begin{tabular}{lcc}
\hline
Property & And V & And VI \\
\hline
$v_r (\kms)$ &   -393.1$\pm4.2$  &-344.8$\pm3.4$ \\
$\sigma_v (\kms)$ & 11.5$^{+5.3}_{-4.4}$ & $9.4^{+3.2}_{-2.4}\kms$\\
$\feh_{spec}$ & -1.6$\pm0.3$ & $-1.3\pm0.14$ \\
$M_{half} (\msun)$ & 2.3$^{+1.5}_{-1.3}\times10^7$ & $2.1^{+1.0}_{-0.8}\times10^7$\\
$[M/L]_{half}$ & 78$^{+51}_{-44}$ &  12.3$^{+5.9}_{-4.7}$ \\
\hline
\end{tabular}
\end{center}
\end{table}

\subsubsection{Masses and dark matter content}

As dSph galaxies are dispersion supported objects, we can use their
internal velocity dispersions to measure a mass for the system, and
infer how dark matter dominated they are. There are several methods in
the literature for this (e.g. \citealt{illingworth76,richstone86}),
however these methods make the assumption that mass follows light,
something we know to be incorrect from measurements of many dSph
systems (e.g. \citealt{walker07,walker09b}, whose high masses cannot
be explained by the luminous matter only. Recent work by
\citet{walker09b} has shown that the mass contained within the
half-light radius ($M_{half}$) of these objects can be reliably
estimated using the following formula:

\begin{equation}
M_{half}=\mu r_{half}\sigma_{v,half}^2
\end{equation}

\noindent where $\mu=580\msun\kpc^{-1}$km$^{-1}$s$^2$, $r_{half}$ is
the half-light radius in kpc and $\sigma_{v,half}$ is the velocity
dispersion within the half-light radius. This simple estimator assumes
that the stellar component is distributed as a Plummer sphere with an
isotropic velocity distribution and a velocity dispersion remains
constant throughout the system. If we apply this estimator to And V
and And VI we obtain the following values for $M_{half}$:
2.3$^{+1.5}_{-1.3}\times10^7\msun$ for And V and
2.1$^{+1.0}_{-0.8}\times10^7\msun$ for And VI, also summarized in
Table~\ref{kprops}. From this, it is trivial to estimate the
mass-to-light ratios for these objects. The $V-$band luminosities for
And V and And VI are $L_V=5.75\times10^6\lsun$ and
$L_V=3.40\times10^7\lsun$, and we obtain the following values:
$[M/L]_{half}$=78$^{+51}_{-44}$ and
$[M/L]_{half}$=12.3$^{+5.9}_{-4.7}$ \citep{mcconnachie06b}. This
demonstrates that each of our objects is likely to be dark matter
dominated.

\subsubsection{Metallicities}

\begin{figure*}
\begin{center}
\includegraphics[angle=0,width=0.45\hsize]{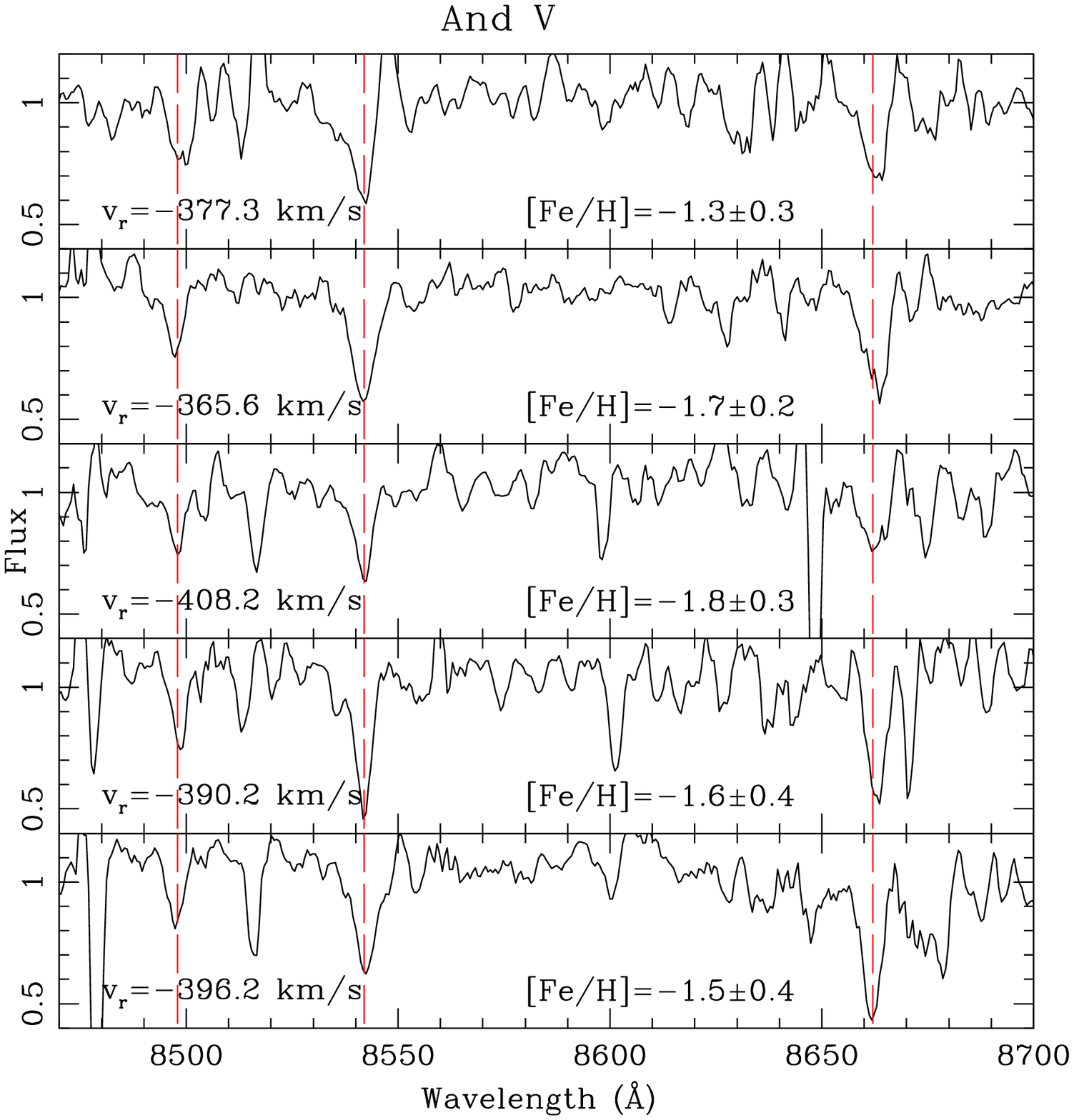}
\includegraphics[angle=0,width=0.45\hsize]{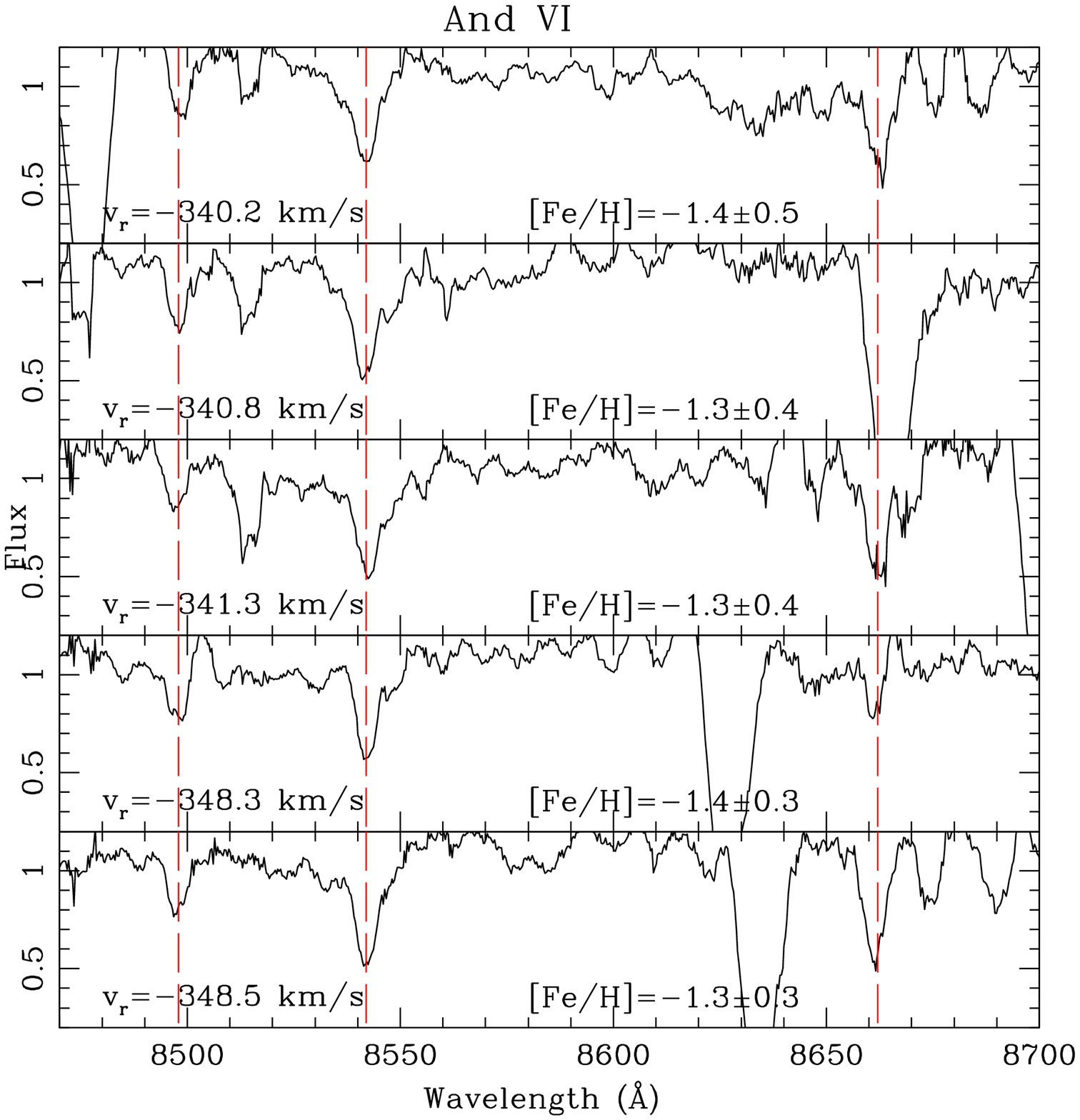}
\caption{5 spectra from individual And V (left) and And VI (right),
  shifted to the rest-frame and normalised in flux. These span a range
  of S:N to demonstrate the quality of our datasets. We indicate the
  positions of the Ca II features with red dashed line. The velocities
  and spectroscopic metallicities of each star are shown also.}
\label{spectra}
\end{center}
\end{figure*}

\begin{figure*}
\begin{center}
\includegraphics[angle=0,width=0.45\hsize]{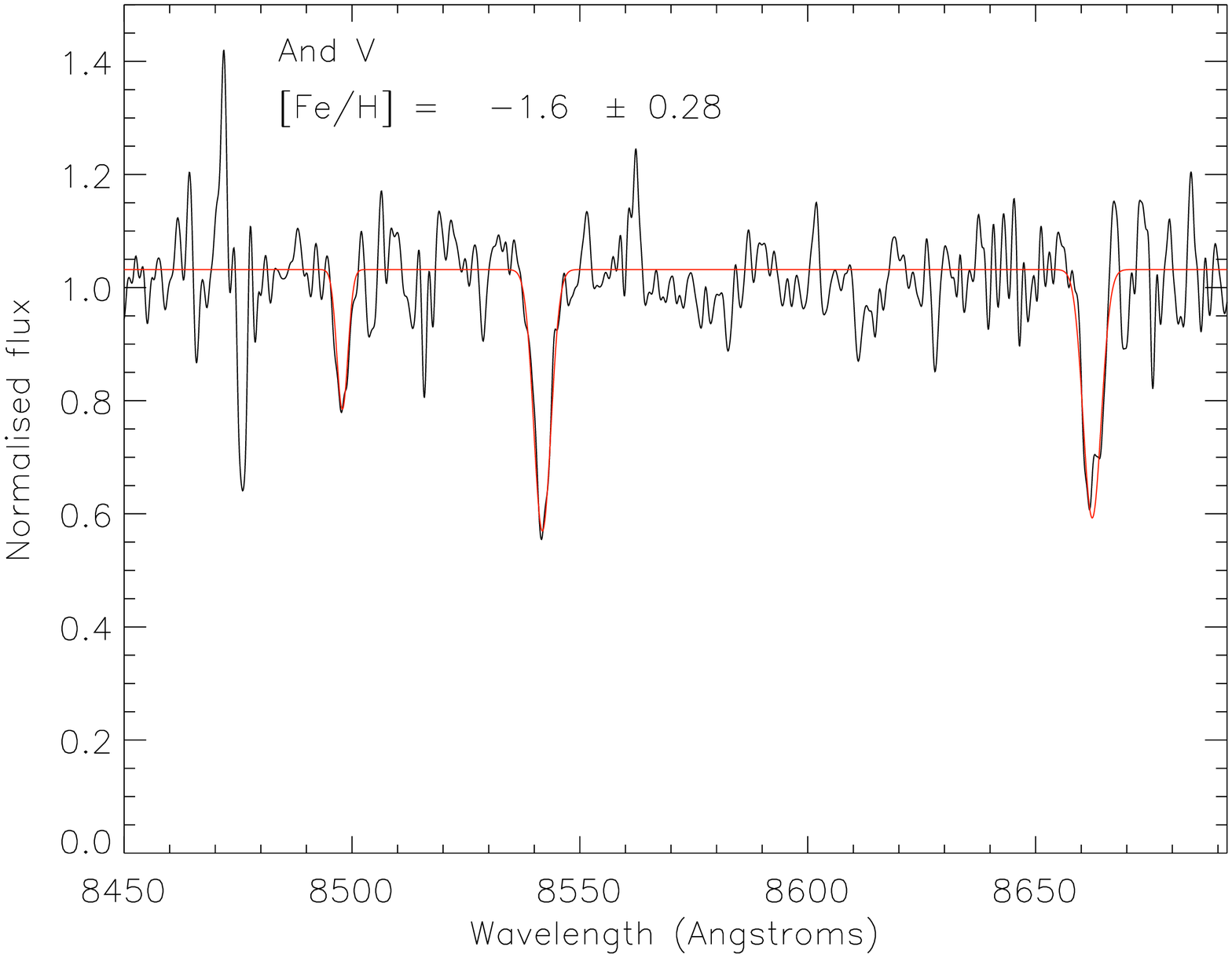}
\includegraphics[angle=0,width=0.45\hsize]{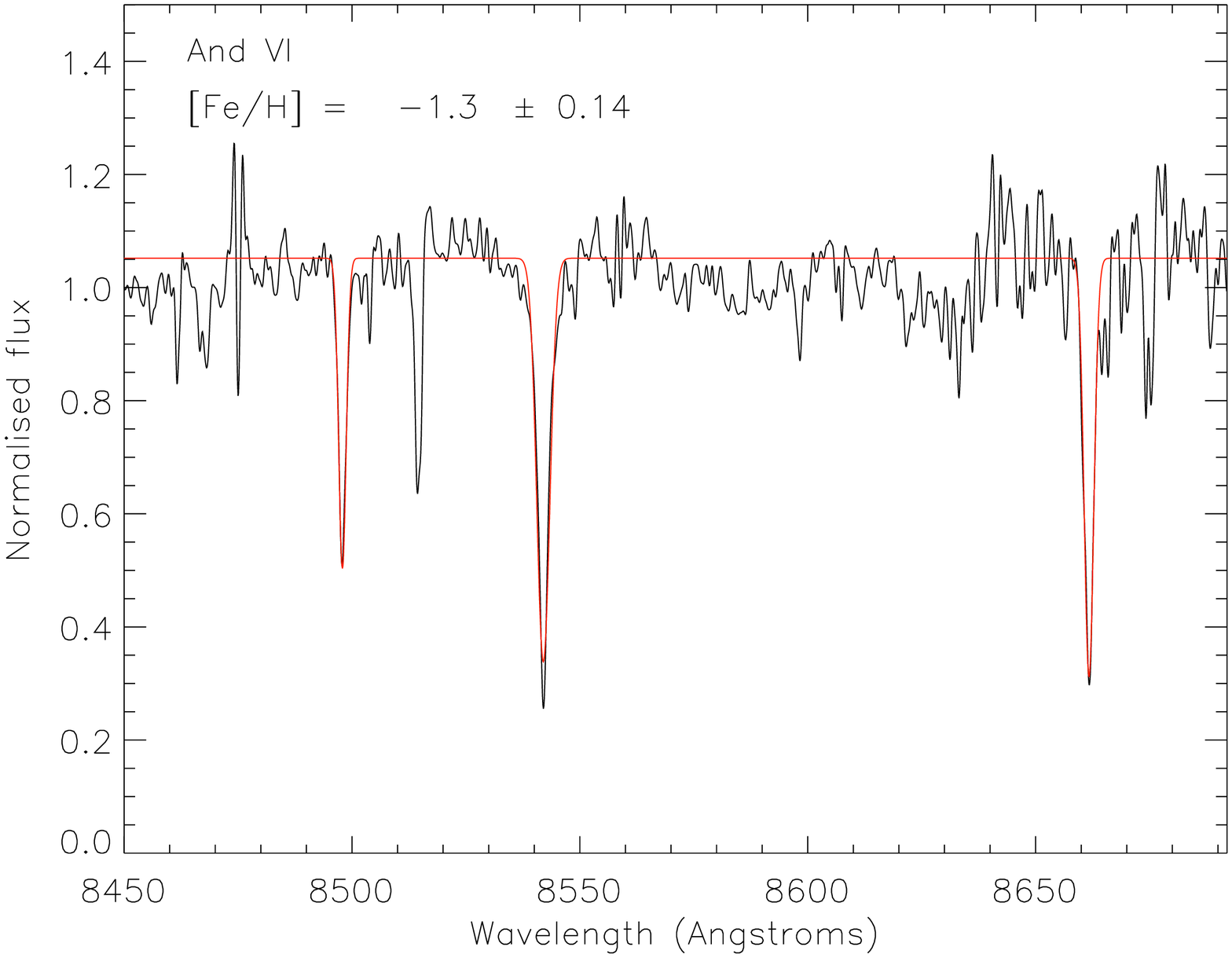}
\caption{Co-added spectra for And V (left) and And VI (right),
  constructed from member stars of each with S:N$>3$. We use the Ca II
  triplet lines to estimate the average metallicity for each dSph and
  find $\feh=-1.6\pm0.3$ and $\feh=-1.3\pm0.14$ for And V and And VI
  respectively. The larger errors in our And V estimates are due to
  the lower resolution of our LRIS data and the fact that we exclude
  the 3rd line of the triplet from our calculation of [Fe/H] as it has
  been artificially broadened by skylines.}
\label{sumspec}
\end{center}
\end{figure*}

In \S~2, we used Subaru photometry to measure the photometric
metallicities of the RGB stars in And V and And VI. Using
\citet{dart08} isochrones with [$\alpha$/Fe]=+0.2 and an age of 10
Gyrs, we deduced average metallicities of [Fe/H]$=-1.8$ and
[Fe/H]$=-1.1$ for And V and And VI respectively. While these values
give us a good sense of the relative metallicities of these objects,
our photometric data do not go deep enough to realise the MSTO of
these dSphs, preventing us from deducing accurate ages for the stellar
populations within. We are also unable to reliably discern their
$\alpha$-abundances, and these factors leave us exposed to the
age--[$\alpha$/Fe]--metallicity degeneracy, which attributes an error
on our estimates of $\pm0.2$ dex. To gain further insight into the
average metallicities of these objects, we can turn to the spectra of
the members of each dSph.

As discussed in \S~2, for both And V and And VI, our observational
setup was such that we observed our stars in the wavelength regime of
the Ca II triplet. The equivalent widths of this strong absorption
feature have been shown by numerous authors
(e.g. \citealt{battaglia08,starkenburg10}) to be a very good proxy for
the measurement of [Fe/H], down to values as low as $\feh=-4$
\citep{starkenburg10}. Thus, we are able to use this feature to
measure the metallicities for the individual stars of And V and
And VI as well as for a composite spectrum for each, where we perform
an error weighted co-addition of the spectra of the confirmed members
of each satellite. To calculate metallicities from the equivalent
widths of the 3 lines (Ca II$_{8498}$,Ca II$_{8542}$,Ca II$_{8662}$)
we use the following relation:

\begin{equation}
\feh=-2.66+0.42[EW-0.64(V_{HB}-V_{RGB})]
\end{equation}

\noindent where
$EW=0.5$~Ca II$_{8498}$+Ca II$_{8542}$+0.6~Ca II$_{8662}$, $V_{HB}$ is
the $V$--band magnitude of the HB at the distance of M31 (25.17,
\citealt{harbeck05}) and $V_{RGB}$ is the $V$--band magnitude of the
star (or weighted average $V$--band magnitude of the sample for the
composite).

In Fig.~\ref{spectra} we show a number of spectra for our And V and
And VI members stars spanning a range of luminosities and S:N, along
with their calculated spectroscopic metallicities. When calculating
errors we not only include the errors on the fit itself, but
systematic errors in measuring the continuum level and the
relationship between the equivalent widths and the measurement of
[Fe/H]. As can be seen, for the star-by-star metallicities our [Fe/H]
measurements carry significant errors (of order 0.3-0.5 dex or
greater), as precision measurements of equivalent widths and the
continuum of each star are difficult to accurately assess in this low
S:N regime. {\bf This inaccuracy can be seen in Tables~\ref{andv}
  and~\ref{andvi}, where the spectroscopic metallicities are often
  very different from the photometrically derived metallicities which
  typically have errors of only $\pm0.2$~dex}. From measuring the
metallicities of the individual members, we find a range of
metallicities for And V from the spectra of [Fe/H]$=-2.2$ --
[Fe/H]$=-1.2$, with a mean of [Fe/H]$=-1.6$. For And VI we find a
range of [Fe/H]$=-0.9$ -- [Fe/H]$=-1.8$ with a mean of
[Fe/H]$=-1.4$. With the co-added spectra, in Fig.~\ref{sumspec}, we
derive metallicities of $\feh=-1.6\pm0.3$ for And V and
$\feh=-1.3\pm0.15$ for And VI. The larger errors on the [Fe/H] for And
V are the result of a number of factors. First, we have a lower number
of member stars for this dSph, meaning the noise in the resulting
spectrum is greater. Second, the spectroscopic resolution of our LRIS
setup is lower than that of our DEIMOS setup, which can cause
broadening effects in the Ca II lines. Finally, as can be seen in both
the individual and combined spectra for this object, the 3rd line of
the Ca II triplet has been broadened significantly by OH absorption
lines in this regime, and so we have excluded this line from our
metallicity calculations, using only the first and second lines. This
is done with the knowledge that the ratio of equivalent widths of the
Ca II lines when they are not saturated should be 0.4:1:0.75
\citep{starkenburg10}, so we can alter the ratios of the Ca
II$_{8498}$ lines used in eqn. 3 from 0.5 to 1.5. We also see
broadening in a number of the And VI spectra (as can be seen in
Fig~\ref{spectra}), but it is less prevalent throughout the sample. In
the case of both And V and And VI, these spectroscopic values agree
very well with those we derived from the photometry. They also conform
to the observed trend of decreasing metallicity with decreasing
luminosity seen in both the MW and M31 dSph populations
\citep{kirby08,kirby11}.

\begin{figure*}
\begin{center}
\includegraphics[angle=0,width=0.45\hsize]{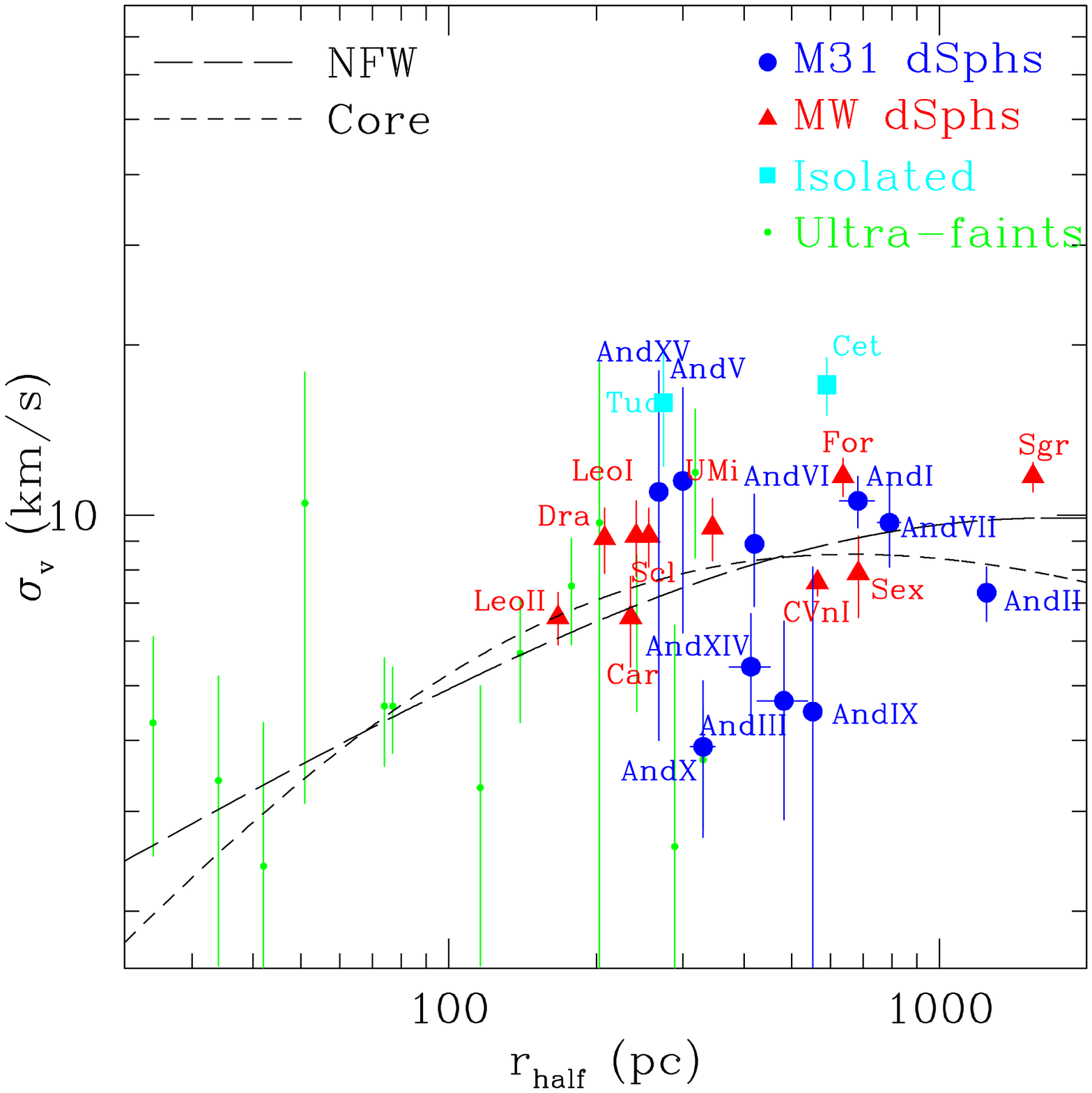}
\includegraphics[angle=0,width=0.45\hsize]{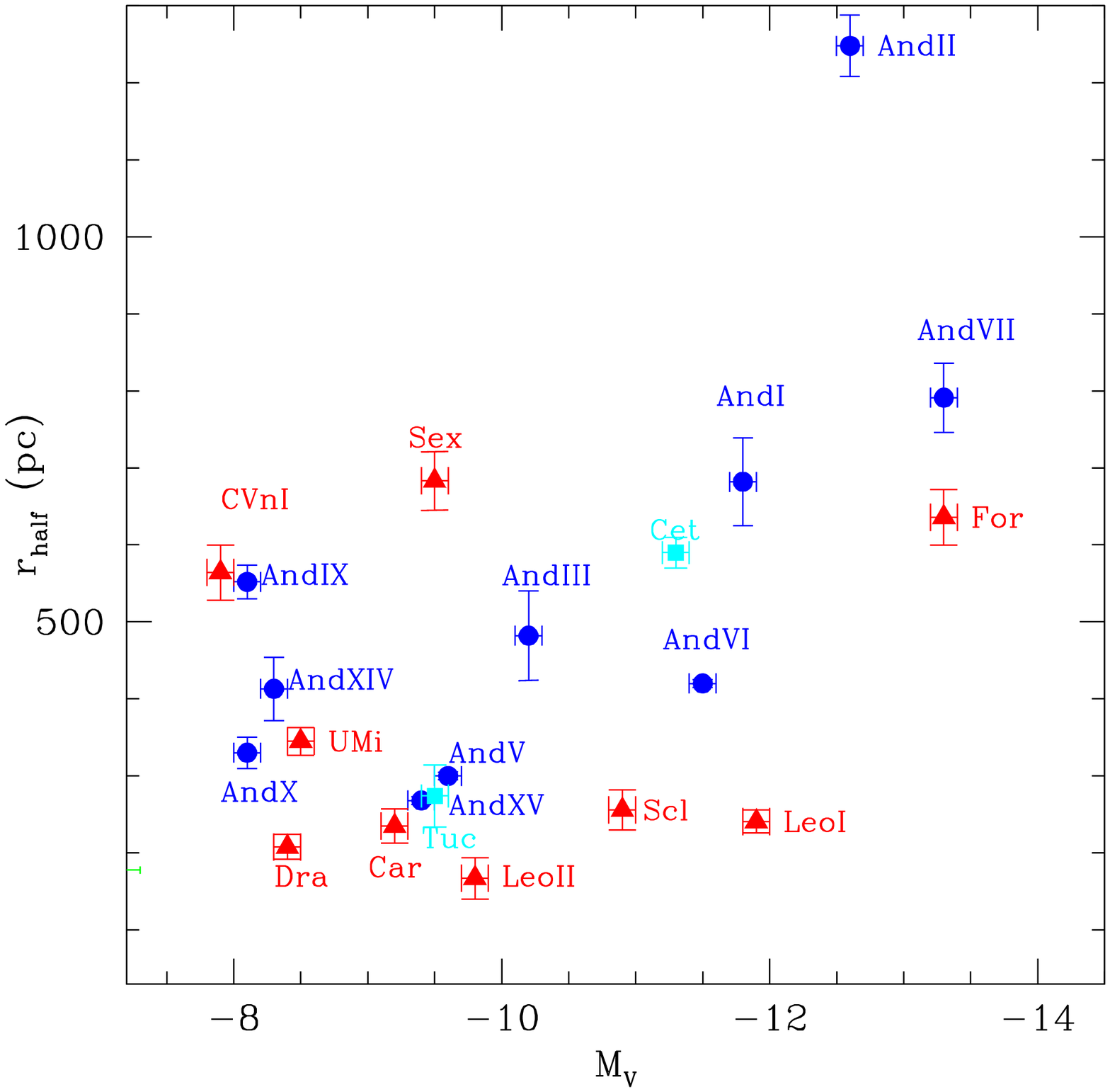}
\includegraphics[angle=0,width=0.45\hsize]{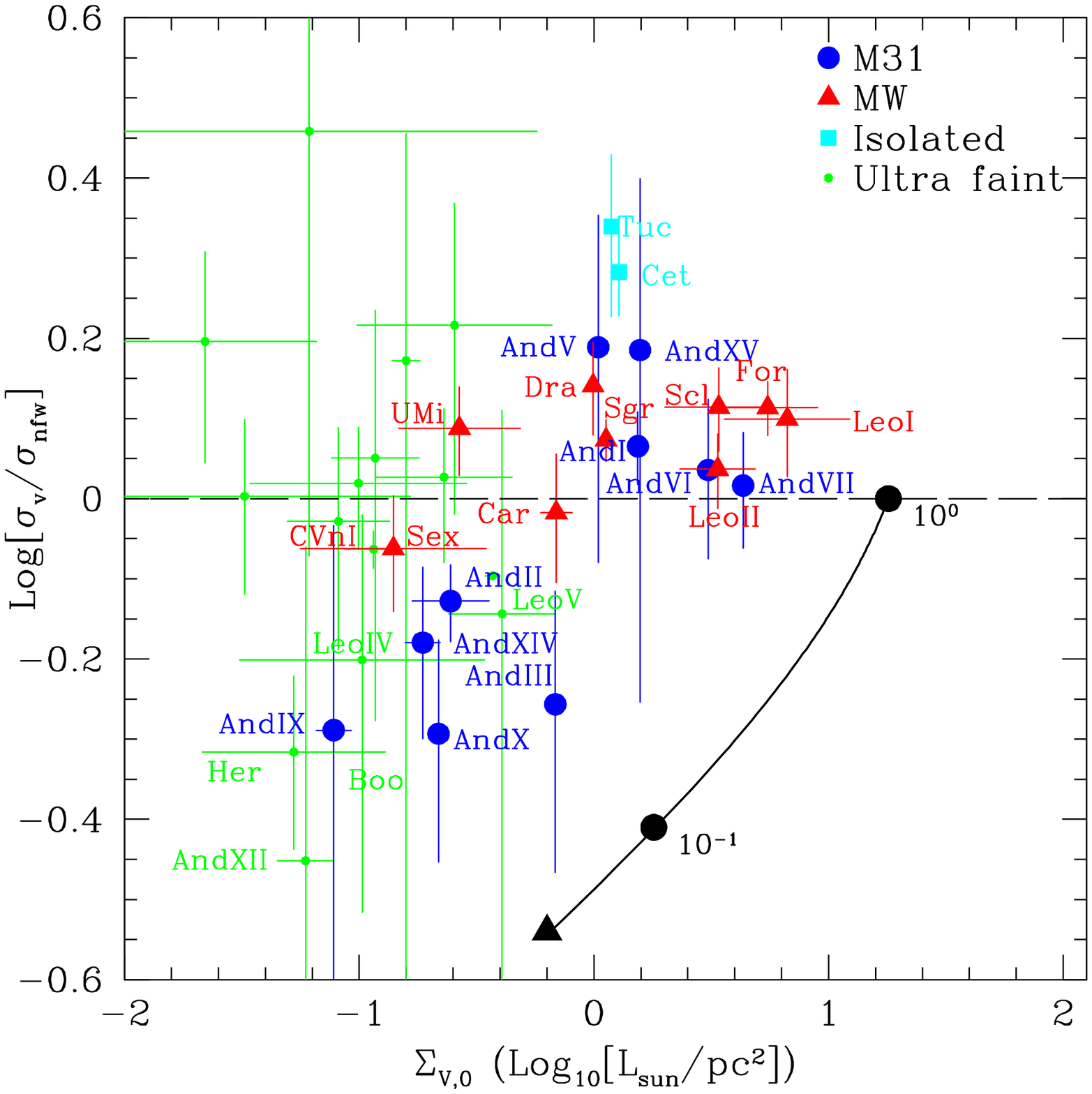}
\caption{{\bf Left: }$r_{half}$ vs. $\sigma$ for the M31 (blue
  circles), MW (red triangles) and isolated (cyan squares)
  ``classical'' dSphs. M31 dSphs tend towards colder values than the
  best-fit MW relations (NFW, Cored and power law shown as long-dash,
  short-dash and solid lines) taken from \citet{walker09b} is observed
  in $\sim50$\% of this sample. The ``classical'' MW dSphs have a
  tendency to scatter above this relation (with the notable exception
  of CVnI). The two isolated dSphs, Cetus and Tucana are
  significantly hotter than the MW mass profiles. We also show the MW
  and M31 ultra-faint objects (i.e. $M_V>-7.9$) as small green
  points. The majority of these are consistent with the relations,
  barring And XII and Hercules which are both colder. {\bf Right:
  }$M_V$ vs. $r_{half}$ for the ``classical'' population, colour-coded
  as before. A number of the M31 dSphs are more
  extended for a given luminosity than their MW
  counterparts. Interestingly, a number of the more extended M31
  objects also fall within the colder regime of the previous
  plot. {\bf Lower: }Central surface brightness ($\Sigma_{V,0}$)
  vs. scatter for all dwarfs about the best fit NFW profile from the
  left hand panel, where the scatter is measured as the difference
  between the observed velocity dispersion from the observed velocity
  dispersion, divided by the expected dispersion in the NFW model,
  $\sigma_v/\sigma_{nfw}$. If these dSphs all inhabited a
  ``Universal'' halo, you would expect to see the points scattering
  around zero, but instead we see a trend of increasingly negative
  scatter with decreasing central surface brightness. We over-plot the
  expected path for an object undergoing tidal disruption to follow in
  this plane, from\citet{penarrubia08a}. The black points represent
  a dwarf which has yet to lose any mass from tides, and the point at
  which it has lost nine tenths of its mass. It follows our observed
  trend well, and we argue that tides have likely played some part in
  the scatter about a ``Universal'' mass profile.}
\label{compare}
\end{center}
\end{figure*}

\section{Comparing the ``classical'' dSphs of M31 and the MW}

In the literature, the term ``classical'' has been applied to the MW
dSph galaxies that were discovered pre--2005 i.e. Carina (Car), Draco
(Dra), Fornax (For), Leo I (LeoI), Leo II (LeoII), Sculptor (Scl), and
Ursa Minor (UMi). All of these dSphs have absolute magnitudes of
$M_V\le-8$. By limiting ourselves to this brighter regime, we can
ensure that we have robust estimates for the parameters we are
interested in, particularly the velocity dispersions. While there are
a few dSphs in M31 with $M_V>-8$ that have been spectroscopically
surveyed (i.e. And XI, XII and XIII, \citealt{chapman07,collins10}),
their RGBs are sparsely populated, leaving us with only a handful of
stars (4--8) with which to determine kinematic properties. This
results in large associated errors. For the ``classical'' dSphs, one
is able to obtain velocities for much larger numbers of members,
giving us velocity dispersions that are much better
constrained. Therefore for the purpose of this study, we require
``classical'' dwarfs to be any dSph with $M_V\le-7.9$. With this
luminosity constraint, we ensure a quality control for our M31 dSphs,
and avoid the regime of the ultra-faints in the MW where a number of
factors (as discussed in the introduction) make the determination of
the masses and velocity dispersions in these systems very
difficult. Multiple epoch data is required for these objects in order
to constrain their true velocity dispersions \citep{koposov11}. We
choose a lower cut of $M_V\le-7.9$ rather than $M_V\le-8$ as it allows
us to include the MW object Canes Venatici I (CVnI) in our
sample. This object has more in common with the classical MW dSphs
than the ultra-faint population, and it is quite similar to a number
of the classical M31 dSphs in terms of its luminous component, with a
larger $r_{half}$ for its luminosity than typical MW objects. Therefore
it is interesting to examine how it compares with both populations.

With the sample defined, we now investigate the claims of
\citet{walker09b} that the MW dSphs are consistent with being born in
a ``Universal'' dark matter halo, i.e. a halo where the mass contained
within a given radius is identical for all dSphs, irrespective of
their luminosity. We plot $r_{half}$ vs. $\sigma_v$, for our sample,
adding our newly derived velocity dispersions for And V and And VI in
Fig.~\ref{compare}, where the MW dSphs are plotted as red triangles,
and the M31 objects as blue circles. Here we overlay the best-fit
cored and Navarro-Frenk-White (NFW) mass profiles from
\citet{walker09b}, which were derived using the $\sigma_v$ and
$r_{half}$ data from MW satellites. Our MW data comes from
\citet{walker09b}, and references within, with an updated value for
the velocity dispersion of Bo\"otes from \citep{koposov11}. Our M31
data are taken from \citet{letarte09}, \citet{collins10},
\citet{kalirai10} and references within. Our two new measurements for
And V and And VI agree well with the \citet{walker09b} relations, as
do And I, And VII and And XV, demonstrating that while M31 does appear
to have a population of colder outliers, among its brighter population
there is a significant fraction of dSphs that agree well with their MW
counterparts. This raises the question, what is the difference between
these systems and their colder brethren? 

Whilst the M31 dSphs scatter towards colder velocity dispersion for a
given size, the MW dSphs scatter in the opposite direction, with 4 of
the 9 ( Dra, Scl, UMi and For) satellites lying over $1\sigma$ from
the NFW and Core profiles in the hotter sense, while only one member
of this population (CVnI) falls below it. Even as we go to
lower luminosities, shown as small green points, we find
only one other dSph that is colder than these relations, the Hercules
dSph, which several authors (e.g. \citealt{sand09,jin10}) claimed may
not be a dSph at all, but a stellar stream of debris formed from the
disruption of a dwarf satellite. Similarly, while 5 of our 10 M31
objects are significantly colder than these relations, only one of
these objects (AndI) appears to be significantly hotter than
predicted. We show the two isolated Local Group dSphs, Cetus
(Cet) and Tucana (Tuc) in Fig.~\ref{compare}, and these are also seen
to be outliers to the MW relations, with hotter dispersions than
expected from the best--fit profiles.

To see how significant these deviations are, we measure the
statistical offset of the ``classical'' dSphs of both the MW and M31
from these profiles. To do this, we calculate the difference between
the velocity dispersion of each object, and the expected velocity
dispersion given by the best--fit NFW and Core profiles for a dSph of
the same half--light radius, and we divide this by the $1\sigma$
errors associated with the object in question. This will allow us to
measure the scatter for both populations, as well as the average
offset, $\langle|\sigma|\rangle$. Taking the MW sample first, we
measure an average deviation of $\langle|\sigma_{MW,NFW}|\rangle=1.64$
and $\langle|\sigma_{MW,Core}|\rangle=1.44$. If these profiles were
truly a good fit for this population, one would expect to see a
scatter of order $\langle|\sigma|\rangle\approx0.2-0.3$. This implies
that the MW dSphs in our sample are not well fit by these mass
profiles. In fact, if we look at the distribution of the scatter, we
find that 5 out of 9, or 55\% of the dSphs in this sample are distinct
from these profiles beyond their 1$\sigma$ errors. Further, we note
that 2 of the MW dSphs, For and CVnI are outliers at $>3\sigma$ for
both profiles. If these profiles were truly Universal, one would not
expect to see two such significant outliers from a population of 9
objects (i.e. 22\%).We also perform this exercise in a sign-dependent
fashion to see if there is a preferred direction for the scatter in
the MW dSphs about these profiles. Here we find
$\langle\sigma_{MW,NFW}\rangle=+0.72\pm0.2$ and
$\langle\sigma_{MW,Core}\rangle=+0.5\pm0.1$, where the errors
represent the standard deviation of these measurements. This suggests
that the MW sample scatter more in the positive or ``hotter''
direction about these best--fit profiles. Performing the same analysis
with the M31 ``classical'' dSphs, we find
$\langle|\sigma_{MW,NFW}|\rangle=1.52$ and
$\langle|\sigma_{MW,Core}|\rangle=1.4$. This shows that the of scatter
about these profiles for the MW and M31 dSphs is very similar, and
that the profiles of \citet{walker09b} are not a demonstrably worse
fit to these objects. The typical errors on the M31 velocity
dispersions are larger than their MW counterparts, and as these are
reduced in size with future observations, we will see if this fit gets
better or worse. Again, we note that there are 2 significant outliers
in the M31 system, And II and And X, both of whom sit more than
$3\sigma$ below the best fit relations, and that 6 of the 10 M31 dSphs
(60\%) are outliers to the relation at $>1\sigma$. Again, assessing
the scatter of the population in a sign--dependent manner, we find
$\langle\sigma_{M31,NFW}\rangle=-0.80\pm0.2$ and
$\langle\sigma_{M31,Core}\rangle=-0.59\pm0.4$ showing that these
objects preferentially scatter in the more negative or ``colder''
direction.  Overall, we find that the distribution about these
profiles (for both M31 and MW) demonstrates that there is more scatter
overall than you would expect from solely their individual errors, as
only 8 out of 19 (or 42\%) objects sit within $1\sigma$ of their
expected values. Finally, we note that the two isolated dSphs, Cetus
and Tucana, are significant ($>4\sigma$ for Cetus and $>2\sigma$ for
Tucana) outliers to the relations.

This spread in the overall population (MW, M31 and isolated dSphs
combined) suggests that some physical factor is driving the scatter
about these relations, and we argue that tidal forces exerted by the
host could play a part in creating the observed differences in both
the central velocity dispersions and surface brightnesses between MW
and M31 satellites. The work of \citet{penarrubia10} explored the
masses and kinematics of the dSph populations for two host galaxies
with the same total mass, but with stellar discs whose masses differed
by a factor of two. They found that the dSphs orbiting the heavier
disc had lower masses and intrinsic velocity dispersions for a given
half light radius than those in the lower disc-mass system, and that
these differences were caused by tidal forces exerted from this more
massive central disc. This mechanism was also discussed in
\citet{walker10} as a possible explanation of the discrepancy between
the M31 dSphs and the \citet{mcgaugh07} mean rotation curve. We know
from various studies (e.g. \citealt{hammer07}) that the stellar disc
of M31 is roughly twice as massive as the MW disc, and we do indeed
see colder dispersions for a given half-light radius in M31 cf. the
MW. It is thus plausible that this larger disc-to-halo mass ratio has
influenced the kinematic properties for a number of M31 dSphs. Those
that seem more typical in terms of their velocity dispersions and
central masses could then either be on orbits about their host that
cause them to feel the tidal forces of the disc less keenly than their
counterparts, such as And V and And VI, or could have been accreted to
M31 at a later time. The tendency for the MW objects to spread in the
opposite direction could also be explained using similar
arguments. Tides may have played a more significant role in the
evolution of the colder CVnI, the only classical MW that falls
significantly below the Walker 2009 relations. While no extreme
extra-tidal population has been reported for this object, it is quite
extended along its major axis (an extension of $\sim2\kpc$,
\citealt{martin08b}).

Another discrepancy between brighter MW and M31 dSphs is that the
latter have a tendency to be more extended for a given luminosity than
their MW counterparts \citep{mcconnachie06b}. In the right hand panel of
Fig.~\ref{compare}, we plot the absolute magnitudes of both the
``classical'' M31 and MW dSph as a function of their half--light
radius. Interestingly, a number of the more extended M31 objects in
this panel also fall within the colder regime of the previous plot,
meaning the colder objects tend to have larger scale radii than those
that are observed to be more typical or hotter. In the lower
panel of Fig.~\ref{compare} we demonstrate this trend by plotting the
logarithm of central surface brightness, $\rm{log}[\Sigma_{V,0}]$
(where we measure $\Sigma_{V,0}=L_{half}/\pi r_{half}^2$) against the
logarithm of the scatter about the best fit NFW profile in the
$\sigma_v-r_{half}$ plane, where the scatter is measured as the ratio
of the observed velocity dispersion to the expected dispersion in the
NFW model, $\rm{log}[\sigma_v/\sigma_{nfw}]$. If these dSphs all
inhabited a ``Universal'' halo, you would expect to see the points
scattering around zero, but instead we see a trend of increasingly
negative scatter with decreasing central surface brightness. We argue
that this is likely driven in part by tidal forces. As the dark matter
halos are gradually stripped of mass, they become less dense, causing
a drop in their velocity dispersions. At the same time, the stellar
surface density decreases as stars are gradually removed. To
illustrate the expected path taken by dSphs undergoing tidal
disruption in this parameter space, we over plot the tidal tracks
generated from eqn. 7 and Table 2 of \citet{penarrubia08a} and note
that it follows the same direction as our data, indicating that tides
have likely played some part in the evolution of these objects away
from their initial mass profiles. The black points represent a dwarf
which has yet to lose any mass from tides, and the point at which it
has lost nine tenths of its mass. This figure clearly demonstrates
that for some of the highest surface brightness MW objects (e.g. Leo
I, Scl and For), the NFW profile from \citet{walker09b} agrees poorly
with the observed dispersions for these objects, underestimating them
by $\sim0.1$ dex. These objects would have their central dispersions
and masses modeled better with halo profiles that possessed a higher
circular velocity than that used by \citet{walker09b} ($V_c\sim18\kms
cf. V_c\sim15\kms$), although such a profile would be demonstrably
worse for the colder MW and M31 objects. This finding strongly argues
against the notion that all dSphs currently reside in a ``Universal''
halo.  We also plot the positions of the ultra-faint dSphs in this
plane, however owing to the large uncertainties in measured values of
luminosity, half--light radius and velocity dispersion, they
demonstrate a large scatter about the expected value of zero. We note
that there are a number of ultra--faints that have significantly
colder dispersions than expected for their surface brightness,
including Hercules, Bo\"otes and And XII, suggesting that tides likely
play an important role in shaping these objects.

Exploring this tidal origin as an explanation for
these differences we turn to the kinematic properties of the two
isolated Local Group dSphs, Cetus and Tucana. Owing to their large
distances from either the MW or M31 (755 kpc and 890 kpc), they are
not thought to have felt a strong tidal force from either galaxy over
the course of their evolution. For the Cetus dSph, analysis of both
INT photometry and Keck DEIMOS spectroscopy
\citep{mcconnachie06b,lewis07} has shown no evidence for previous or
ongoing tidal disruption, and its large tidal radius (6.6 kpc
\citealt{mcconnachie06b}) implies that tidal forces from either the MW
or M31 are unlikely to have played a significant role in its
evolution. Similarly, Tucana shows no obvious signs of tidal
disruption, other than an absence of HI gas \citep{fraternali09} which
could have been expelled by stellar feedback, or tidally removed. As
it has an unusually large receding velocity with respect to the MW
($v_r=+98.9\kms$, \citealt{fraternali09}), it has been argued that
Tucana may be on a highly elongated orbit about the MW, bringing it in
close proximity to our Galaxy $\sim10$Gyrs ago, meaning it may have
experienced stronger tidal forces than Cetus. Given its current
position, it is unlikely to have completed more than one orbit within
the Local Group, and therefore has experienced less tidal-stirring
over the course of its history compared with the M31 and MW dSphs. In
our analysis here, we find both these objects are positioned in the
hotter $r_{half}-\sigma_v$ regime, with Cetus being a more extreme
outlier than Tucana. This fits with our expectation if tidal forces
from the host are the cause of this scatter about a Universal profile.

These results demonstrate that, while other factors may also be
involved in shaping the physical and dynamical properties of the dark
matter halos of dSphs, the effects of tidal forces exerted by the host
galaxy play an important role in setting the underlying mass profiles
of its associated satellite population. In the context of the MW and
M31 systems, we argue that the observed scatter towards colder
velocity dispersions in the M31 subhalos is likely due to the larger
disc-to-halo mass ratio in this system compared with the MW.

\section{Conclusions}

In this study, we set out to further investigate the results of
\citet{collins10} and \citet{kalirai10} who noted that the velocity
dispersions and central masses of a number of the M31 dSphs differed
significantly from MW dSphs of similar scale-radii, making them
outliers to the universal mass profiles derived for the MW dSphs, and
the \citet{mcgaugh07} mean rotation curve
\citep{walker09b,walker10}. We restricted ourselves to studying the
``classical'' dSphs in both galaxies, where classical was defined as
any dSph with $M_V<-7.9$ to avoid complications introduced by the
uncertain nature of some ultra-faint dwarfs, and low number statistics
for the faint M31 dSphs. We also presented updated structural
properties and kinematic properties for two of the classical M31
population, And V and And VI, for whom only systemic velocities have
previously been reported \citep{guhathakurta00}. We updated values for
their half--light and tidal radii to $r_{half}=292\pm22$pc and
$r_{half}=440\pm16$pc, $r_t=1.2\pm0.2\kpc$ and $r_t=1.6\pm0.2\kpc$,
their PA ($\theta=32\pm2\deg$ and $\theta=164\pm2\deg$) and
ellipticities ($\epsilon=0.17\pm0.02$ and $\epsilon=0.39\pm0.02$) for
And V and And VI respectively. In terms of their kinematics, we
measured systemic velocities of $v_r=-393.1\pm4.2\kms$ and
$344.8\pm2.5\kms$, and dispersions of
$\sigma_v=11.5^{+5.3}_{-4.4}\kms$ and $\sigma_v=9.4^{+3.2}_{-2.4}\kms$
for And V and And VI respectively, meaning that these two dSph appear
very typical for objects of their size when compared with MW
dSphs. This result shows that not all of the M31 dSphs reside in
significantly different dark matter halos to those of the MW. When
assessing the classical MW and M31 in the $r_{half}-\sigma_v$ plane,
we find that with respect to the best--fit mass profiles of
\citet{walker09b}, scatter about these is observed in both the
positive and negative directions, and is greater than would be
expected from the measurement errors alone. The scatter in the
positive or ``hotter'' direction occurs predominantly within the MW
population with only one M31 object (AndI) found to scatter in this
direction, and similarly, all but one of the ``colder'' objects are
M31 dSphs (the exception being CVnI). We also find a tendency for the
``hotter'' dSphs to be more compact for a given luminosity than the
``colder'' objects. This is seen when comparing the central
surface brightnesses of these objects with deviations of their
velocity dispersions from the best--fit \citet{walker09b} NFW
profile. In this plane, we can clearly see that deviations towards
colder velocity dispersions increase as surface brightness
decreases. Analysing this in the framework of \citet{penarrubia10}
where the colder velocity dispersions of the M31 dSphs are suggested
to be a result of a more massive disc-to-halo mass ratio in M31
compared to the MW, we argue that the underlying mass profiles for
dwarf galaxies are not Universal, and are influenced the baryonic
component of disc galaxies, causing variation from host to host.\\ \\

\noindent {\bf ACKNOWLEDGEMENTS}\\

\noindent M. L. M. Collins would like to acknowledge support from the
UK Science and Technology Funding Council (STFC) for the provision of
her PhD. stipend. We also thank the referee for their helpful comments
and suggestions for this manuscript. \\

\noindent The data presented herein were obtained at the W.M. Keck Observatory,
which is operated as a scientific partnership among the California
Institute of Technology, the University of California and the National
Aeronautics and Space Administration. The Observatory was made
possible by the generous financial support of the W.M. Keck
Foundation. Based in part on data collected at Subaru Telescope, which is operated
by the National Astronomical Observatory of Japan.
\bibliography{mnemonic,michelle}{} \bibliographystyle{mn2e}

\end{document}